\documentclass[10pt,a4paper]{article}

\newcount\driver
\newcount\bozza

\font\msytw=msbm10 scaled\magstep1%
\font\msytww=msbm8 scaled\magstep1%
\font\msytwww=msbm7 scaled\magstep1%
\font\indbf=cmbx10 scaled\magstep2%

{\count255=\time\divide\count255 by 60 \xdef\hourmin{\number\count255}
        \multiply\count255 by-60\advance\count255 by\time
   \xdef\hourmin{\hourmin:\ifnum\count255<10 0\fi\the\count255}}

\let\a=\alpha \let\b=\beta  \let\g=\gamma     \let\d=\delta  \let\e=\varepsilon
\let\z=\zeta  \let\h=\eta    \def\th{\theta}
\let\k=\kappa   \let\l=\lambda
\let\m=\mu    \let\n=\nu    \let\x=\xi        \let\p=\pi      \let\r=\rho
\let\s=\sigma \let\t=\tau        
   \let\o=\omega 
\let\G=\Gamma \let\D=\Delta     \let\L=\Lambda  
           
\let\O=\Omega \let\Y=\Upsilon

\def\\{\hfill\break} \let\==\equiv

\let\io=\infty 

\let\0=\noindent

\def\ie{\hbox{\it i.e.\ }}\def\eg{\hbox{\it e.g.\ }}
\def\der{\hbox{\rm d}}
\let\dpr=\partial 

\def\defin{{\buildrel def\over=}}
\def\tende#1{\,\vtop{\ialign{##\crcr\rightarrowfill\crcr
 \noalign{\kern-1pt\nointerlineskip}
 \hskip3.pt${\scriptstyle #1}$\hskip3.pt\crcr}}\,}
\def\otto{\,{\kern-1.truept\leftarrow\kern-5.truept\to\kern-1.truept}\,}

\def\MM{{\cal M}}
\def\CC{{\cal C}}
\def\TT{{\cal T}}\def\MM{{\cal M}}
\def\LL{{\cal L}}
\def\DD{{\cal D}}\def\GG{{\cal G}}
\def\OO{{\cal O}}\def\QQ{{\cal Q}}\def\XX{{\cal X}}

\def\Im{{\rm Im}\,}
\def\Tr{{\rm Tr}\,}
\def\lis{\overline}

\def\indica{\leaders \hbox to 0.5cm{\hss.\hss}\hfill}
\def\guida{\leaders\hbox to 1em{\hss.\hss}\hfill}
\mathchardef\oo= "0521
\def\SRB{\mbox{\tiny\bfseries SRB}}

\def\xx{{\bf x}}
\def\yy{{\bf y}}

\def\Halmos{\hfill\vrule height6pt width4pt depth2pt \par\hbox to \hsize{}}
\def\virg{\quad,\quad}

\def\oo{{\underline \omega}}

\def\qed{\raise1pt\hbox{\vrule height5pt width5pt depth0pt}}

\def\Val{{\rm Val}}
\def\indic{\hbox{\raise-2pt \hbox{\indbf 1}}}

\def\RRR{\hbox{\msytw R}} 
 \def\CCC{\hbox{\msytw C}}
 
\def\NNN{\hbox{\msytw N}} 
 \def\ZZZ{\hbox{\msytw Z}}
\def\zzzz{\hbox{\msytww Z}} \def\zzz{\hbox{\msytwww Z}}
\def\TTT{\hbox{\msytw T}}

\def\ins#1#2#3{\vbox to0pt{\kern-#2 \hbox{\kern#1 #3}\vss}\nointerlineskip}
\newdimen\xshift \newdimen\xwidth \newdimen\yshift
\newcount\griglia

\def\insertplot#1#2#3#4#5#6{%
\begin{figure}[h]
\begin{center}
\vspace{#2pt}
\begin{minipage}{#1pt}
#3
\ifnum\driver=1
\griglia=#6
\ifnum\griglia=1
\openout13=griglia.ps
\write13{gsave .2 setlinewidth}
\write13{0 10 #1 {dup 0 moveto #2 lineto } for}
\write13{0 10 #2 {dup 0 exch moveto #1 exch lineto } for}
\write13{stroke}
\write13{.5 setlinewidth}
\write13{0 50 #1 {dup 0 moveto #2 lineto } for}
\write13{0 50 #2 {dup 0 exch moveto #1 exch lineto } for}
\write13{stroke grestore}
\closeout13
\includegraphics{griglia.ps}\fi
\includegraphics{#4.pst}\fi
\ifnum\driver=2
\fi
\end{minipage}
\end{center}
\caption{#5}
\end{figure}
}

\newdimen\shift \shift=-1truecm
\def\lb#1{%
\ifnum\bozza=1
\label{#1}\rlap{\kern\shift{$\scriptstyle#1$}}
\else\label{#1}
\fi}

\def\be{\begin{equation}}
\def\ee{\end{equation}}
\def\bea{\begin{eqnarray}}\def\eea{\end{eqnarray}}
\def\bean{\begin{eqnarray*}}\def\eean{\end{eqnarray*}}
\def\bfr{\begin{flushright}}\def\efr{\end{flushright}}
\def\bc{\begin{center}}\def\ec{\end{center}}
\def\ba#1{\begin{array}{#1}} \def\ea{\end{array}}
\def\bd{\begin{description}}\def\ed{\end{description}}
\def\bv{\begin{verbatim}}\def\ev{\end{verbatim}}
\def\nn{\nonumber}
\def\Halmos{\hfill\vrule height10pt width4pt depth2pt \par\hbox to \hsize{}}
\def\pref#1{(\ref{#1})}
 
\def\virg{\quad,\quad}

\def\bye{\end{document}}

\title{\bf Analyticity of the SRB measure of a lattice of  coupled Anosov 
diffeomorphisms of the torus} 
\author{F. Bonetto\thanks{School of Mathematics, GeorgiaTech,
        Atlanta, USA} 
       \and P.Falco\thanks{Dipartimento di Matematica, Universit\`a 
        la Sapienza, Roma}\
       \and A.Giuliani\thanks{Dipartimento di Fisica, Universit\`a 
       la Sapienza, Roma; and INFN, sezione di Roma 1}}

\begin{document}
\maketitle

\section*{\hspace*{7.8cm}\normalsize Abstract}
{\small\sl We consider the ``thermodynamic limit'' of a
$d$-dimensional lattice of hyperbolic dynamical systems on the 2-torus,
interacting via weak and nearest neighbor coupling. We prove that the
SRB measure is analytic in the strength of the coupling. The proof is
based on symbolic dynamics techniques that allow us to map the SRB
measure into a Gibbs measure for a spin system on a
$(d+1)$-dimensional lattice. This Gibbs measure can be studied by an
extension ({\it decimation}) of the usual ``cluster expansion''
techniques. }

\newpage
\section{Introduction and Main Results}\lb{s1}\setcounter{equation}{0}

In recent years a lot of attention has been devoted to the relation
between non equilibrium statistical mechanics and dynamical systems
theory. According to the point of view of Ruelle, Cohen and Gallavotti
\cite{Ru78} \cite{GC95} a mechanical system evolving in a steady state
can be described by a hyperbolic dynamical system and its properties
can be deduced from the ``natural'' or SRB distribution (see below for
a precise definition) associated with this dynamical system. This line
of investigation has already produced several interesting results both
analytical, like the ``Fluctuation Theorem'', see \cite{GC95}, or
numerical, like the works of Evans and Morris, see \cite{EM90}, and Moran
and Hoover, see \cite{MH87}. Nonetheless, most of the work has been devoted
to low dimensional dynamical system, due to their accessibility both
to analytical and to numerical study.  In this paper we want to study
the properties of the SRB distribution for a class of simple systems
in very high dimension. For more references on this kind of systems see
\cite{JP98}. The precise model model we study here is taken taken from
\cite{BKL03}.  

We start considering a linear hyperbolic automorphism of the 2-torus
$\TTT^2$. To be definite, we will always consider the so called
{\it Arnold cat map} $s_0:\TTT^2\to\TTT^2$ defined by the action modulus
$2\pi$ of the matrix 
\be 
A=\pmatrix{1 & 1\cr 1 &2 \cr}. \lb{1} 
\ee 
Note that
the matrix $A$ admits two orthogonal eigenvectors $v_\pm$ whose
respective eigenvalues $\l_\pm$ are such that $\l_+>1>\l_-$ and
$\l_+\l_-=1$. For this reason the dynamical system $s_0$ is uniformly
hyperbolic and the stable and unstable manifolds at any point $\phi\in
\TTT^2$ are given by $W^\pm_\phi(t)=\phi+v_\pm t \ {\rm mod} 2\pi$.

>From $s_0$ we can construct the {\it uncoupled} lattice dynamics by
considering as phase space the Cartesian product
$\TT=(\TTT^2)^{\zzz^d}$ (namely any point $\psi\in\TT$ has Cartesian 
components $\{\psi_\x\}_{\x\in\zzz^d}$), 
equipped with the metric
$d(\psi,\psi')=\sum_\x 2^{-|\x|}\hat d(\psi_\x,\psi'_\x)$ where $\hat
d(\psi_\x,\psi'_\x)$ is the usual metric on $\TTT^2$ and
$|\x|=\sum_{i=1}^d |\x_i|$ for $\xi\in\ZZZ^d$. On $\TT$ the map $S_0$
acts simply as
\be 
S^\x_0(\psi)=\big(S_0(\psi)\big)_\x=s_0(\psi_\x). 
\ee
Note that the stable and
unstable manifold of $S_0$ at a point $\psi$ are the Cartesian product
of the stable and unstable manifold of $s_0$ for the points
$\psi_\x\in\TTT^2$, \ie 
$W^\pm_{0,\psi}(\z)=\psi+\sum_\x w_{0,\pm}^{(\x)}\z_\x\ {\rm mod}\ 2\pi$,
where $w_{0,\pm}^{(\x)}$ is the tangent vector to $\TT$
that has null component on the tangent space to every $\TTT^2_\h$ but
for $\TTT^2_\xi$ where it coincides with $v_\pm$. The action of $S_0$
on $W^\pm(\psi)$ is naturally given by a diagonal linear
transformation.

We observe that the special choice of the matrix $A$ plays no role in the
following. Indeed we will show in Appendix A that our results stay
true if we replace $s_0$ with any uniformly hyperbolic analytic automorphism
of $\TTT^2$, not necessarily linear.

To add a coupling to this system we consider an analytic function
$g:\TT\to\TTT^2$ and define 
\be 
S^\x_\e(\psi)\defin s_0(\psi_\x)+\e g(\r^\x\psi)\defin s_0(\psi_\x)+\e f^\x(\psi)
\lb{3}\ee
where $(\rho^\x\psi)_\h=\psi_{\h+\x}$, \ie $\rho$ is the group of the
translations on $\ZZZ^d$. This means that the function $f:\TT\to\TT$,
whose $\x$ component is $f^\x=g\circ\r^\x$,
is translation invariant. We want $f$ to be short ranged: let the 
nearest neighbor sites of the site $\x$ be  
$nn(\x)=\{\h\ :\ |\x-\h|\leq 1\}$, we will assume that $g$ depended only on
$\psi_{nn(0)}$, where we have used the notation $\psi_{V}=\{\psi_\x |
\x\in V\}$ for $V\subset \ZZZ^d$. This implies that $S^\x_\e(\psi)$
depends only on $\psi_{nn(\x)}$. More generally we could have 
assumed that
$g$ depends only on $\psi_V$ where $V$ is any finite subset of
$\ZZZ^d$ containing $0$ but this would  not have changed 
the substance of the
following arguments. Moreover we will take $g$ analytic in all its
arguments.

The dynamical system $S_\e$ admits many invariant measures. Among them
is the ``natural'' or SRB measure defined
as the weak limit of the volume measure on $\TT$ under the evolution
defined by $S_\e$, when such a limit exists and is ergodic. Being
$\TT$ infinite dimensional, to properly define this concept we will
consider finite dimensional approximations. Let $\TT_N=(\TTT^2)^{V_N}$
where $V_N$ is the cube of side $2N+1$ in $\ZZZ^d$ centered at the
origin. To define the restriction of $S_\e$ to $\TT_N$ we have to fix
the boundary conditions: we choose periodic ones. To this extent note
that $\TT_N$ can be naturally identified with the submanifold of $\TT$
formed by the points periodic of period $2N+1$. Moreover $S_\e$ leaves
such a manifold invariant so that we can define
$S_{\e,N}\defin S_\e\big|_{\TT_N}$. If no confusion can arise we will
suppress the index $N$.

We can now define the SRB measure for $S_{\e,N}$ as
$\mu_N^{SRB}=\lim_{T\to\infty}\frac{1}{T}
\sum_{t=0}^{T-1}(S^*_{\e,N})^t\mu_N^0$ where the limit must
be understood as a weak limit and $\mu_N^0=\prod_{\x\in V_N}
d\psi_\x/(2\p)^2$ is the Lebesgue measure on $\TT_N$. The existence of
such a measure follows from rather general theorem on hyperbolic
dynamical systems, if $\e$ is sufficiently small (see for example
\cite{GBG03} and references there). Moreover
$\mu_N^{SRB}$ is ergodic, always for small $\e$, and we have that
$\mu_N^{SRB}(\OO)=\lim_{T\to\infty}\frac{1}{T}
\sum_{t=0}^{T-1}\OO(S_\e^t(\psi))$ for $\mu_N^0$ almost every $\psi$,
where $\OO$ is an {\it observable}, \ie a Hold\"er continuous function from
$\TT_N$ to $\RRR$. This means that $\mu_N^{SRB}$ is the {\it statistic} of $S_\e$. 
It is well known that the SRB measure is still well defined in the limit
$N\to\io$, for $\e$ small enough. This was first proved 
by Bunimovich and Sinai in \cite{BS88}. Starting from this work, the model
eq.\pref{3} (or similar models of coupled expanding
automorphisms of the circle) has been widely 
studied in the literature, see for instance 
\cite{PS91},\cite{BK95},\cite{BK96},\cite{BK97}
\cite{JM95}, \cite{JP98}, \cite{BEIJK98}, \cite{JdL00}. Many properties 
of such systems are well known, mainly uniqueness 
of the SRB state in the thermodynamic limit and exponential
decay of correlations (see \cite{BK95},\cite{BK96},\cite{BK97} for
a proof of these properties).

We further investigate the regularity properties of the
limiting measure.
We show that $\mu_N^{SRB}$ depends analytically on
$\e$. This means that if we consider an analytic observable $\OO$, \ie
an analytic function from $\TT_N$ to $\RRR$, we have that
$\mu_N^{SRB}(\OO)$ is an analytic function in a domain that depends on
the analyticity properties of $\OO$.

The main point of this work is to show that such a property 
remains true when $N\to\infty$, \ie we want to
show that the domain of analyticity of $\mu_N^{SRB}$ does not shrink to
0 when $N\to\infty$. More precisely we say that $\OO:\TT\to\RRR$ is a
{\it local observable} if it depends only on $\psi_V$ 
for some finite
$V\subset \ZZZ^d$. We can summarize our main results as follows 
\\ 
\\
{\bf Theorem}: {\it Given $S_\e$ as above and a local observable $\OO$
we have
\begin{description}
\item{(1)} $\mu^{SRB}(\OO)=\lim_{N\to\infty}\mu_N^{SRB}(\OO)$ exists
  uniformly in $\e$ for $\e<\e_0$ independent on $\OO$.
\item{(2)} if $\OO$ is local and analytic then $\mu^{SRB}(\OO)$ is
  analytic in $\e$ for $\e<\e_0(\OO)$. 
\end{description}
}

The proof is mainly based on the possibility of mapping the SRB
distribution into the Gibbs state of a suitable spin system on
$\ZZZ^{d+1}$ and on the extension of classical techniques used to study
such Gibbs states (\ie
cluster expansion) to the particular ones that occur in our system. 
The key point in order to get analyticity of the measure is proving that 
the SRB potentials (\ie the potentials
of the Gibbs state the SRB measure is mapped into) 
are rapidly decaying. 
Once this decay is proved, analyticity follows
via standard techniques. Analyticity of the measure and convergence
of cluster expansion imply in 
particular uniqueness of the Gibbs measure and exponential
decay in space and time of the correlations of 
H\"older continuous observables, see for instance \cite{GBG03}.
Our proof can also be adapted to the case of coupled analytic expanding
circle map: in fact also these models can be mapped
into spin systems, and proceeding as below one can prove that the
SRB potentials satisfy the same decaying properties.

The rest of the paper paper is organized as follows.
In section 2 we give a brief review of the main properties 
of smooth uniformly hyperbolic systems and we briefly describe the
construction that allows the above quoted mapping. The detailed proofs of this
properties are postponed to sections 3, 4 and 5. Finally in section 6
we complete the proof of the main theorem. Appendix A contain a
direct extension of our results to the case in which the uncoupled
dynamics is not linear.
Appendix B deals with an application. In the contest of the physical
application of dynamical systems (see the beginning of this
introduction) a special status has been given to a particular
observable, the {\it phase space contraction rate} defined as
$\eta_+(\psi)=\log|\det(DS_\e(\psi))|$ where $DS_\e$ is the
differential of $S_\e$. Being our system infinite it is more
interesting to study the {\it local} phase space contraction rate
$\eta_V(\psi)$ defined by taking the determinant of a (large) minor of
$DS_\e$. We show, for a large class of couplings $f$, that $\eta_V$ has a
positive average and that it obeys a large deviation principle, \ie
its large deviation are asymptotically described by a free energy
functional.

\section{Anosov Systems}\lb{s2}\setcounter{equation}{0}

\subsection{Geometric Properties}\lb{ss2.1}

A dynamical system on a smooth compact manifold, whose dynamics is
given by a uniformly hyperbolic invertible map is called an {\it Anosov system}.
>From the general theory we know that Anosov systems are structurally
stable, namely, given two Anosov diffeomorphisms $S$, $S'$ on a
manifold $\O$ that are sufficiently close in the $C^2$ topology,
there exist a {\it conjugation} $H:\O\otto\O$ 
such that $S\circ H=H\circ S'$. 

In our situation this implies the existence of a map
$h_\e:\TT_N\otto\TT_N$ such that  
\be S_\epsilon\circ h_\epsilon =
 h_\epsilon\circ S_0\;,\lb{2.1}\ee
at least if $\e$ is small enough ({\it a priori} not uniformly in
$N$). The first step of our proof consist in showing that $h_\e$ is
analytic in $\e$ uniformly in $N$. More precisely we will construct
$h_\e$ directly for $N=\infty$. Its finite $N$ version can be obtained
by restricting it to $\TT_N$. We note that $h_\e$ is, in general, only
Hold\"er continuous in the variable $\psi$. By this we mean that there
exist constant $c$ and $\beta$ such that
$d(h_\e(\psi),h_\e(\psi'))\leq c d(\psi,\psi')^\beta$. For this reason
we cannot say that the SRB measure of $S_\e$ is just the image under
the map $h_\e$ of the SRB measure for $S_0$, \ie of the Lebesgue
measure on $\TTT^2$. Notwithstanding $h_\e$ will play a crucial role in
the construction on $\mu_N^{SRB}$.

As we saw in the introduction the tangent space ${\rm T}_\psi\TT$ to
$\TT$ on a point $\psi$ can be split in two subspaces $E^+_\psi$ and
$E^-_\psi$ such that ${\rm T}_\psi\TT=E^+_\psi\oplus E^-_\psi$.
Moreover the distributions $E^\pm$ are continuous and invariant under
$S_0$, \ie $\big(DS_0\ E_\psi^\pm\big) =E_{S_0\psi}^+$ and we have
\begin{eqnarray}
&\|DS_0^nw\|\leq C\l_-^n\|w\|\qquad \hbox {for } w\in E^-_\psi\cr
&\|DS_0^{-n}w\|\leq C\l_+^{-n}\|w\|\qquad \hbox {for } w\in E^+_\psi
\end{eqnarray}
$E^+_\psi$ and $E^-_\psi$ are called the {\it stable} and {\it unstable} 
subspaces,
respectively. In the case of $S_0$ all these properties are trivially
true. In particular we can consider on $E^\pm_\psi$ the basis formed
by the vectors $\big\{{w_{0,\pm}^{(\x)}} \big\}_{\x\in \zzz^d}$.

We will show in section 4 that such a splitting can be
constructed also for $S_\e$, again uniformly in $N$, \ie we will
prove the existence of the stable and unstable subspaces
$E^\pm_{\e,\psi}$ for $S_\e$. Moreover we will show that
$E^\pm_{\e,h_\e(\psi)}$ is an analytic function of $\e$, although is
only Hold\"er continuous in $\psi$. This will turn out to be the right
regularity to study the SRB measure. To do this we will directly
construct the vectors of the basis $\big\{ w_{\e,\pm}^{(\x)}(\psi)
\big\}_{\x\in \zzz^d}$ that coincide 
with $ \big\{w_{0,\pm}^{(\x)}\big\}_{\x\in \zzz^d}$ for
$\e=0$. 

\subsection{Symbolic Dynamics}\lb{ss2.2}

The main property that allow to study analytically the SRB measure for
an Anosov map $S$ acting on a manifold $\MM$ is the existence of 
Markov partitions. We call a
collection $\QQ_i$, $i=1,\ldots,n$, of closed subsets of $\MM$ a {\it
partition} if $\cup_i \QQ_i=\MM$ and
$\QQ_i\cap\QQ_j=\partial\QQ_i\cap\partial\QQ_j$ for every
$i\not=j$. For every sequence $\s=\{\s_t\}_{t\in\zzz} \in\{1,\ldots,n
\}^{\zzz}$ we can
define the set $\XX(\s)=\cap_{t=-\infty}^\infty
S^t(\QQ_{\s_t})$. Due to the hyperbolicity properties of $S$ if
$\QQ_i$ are small enough $\XX(\s)$ contains at most one point. This
allow us to construct a {\it symbolic dynamics} \ie a map from a
subset $\Sigma$ of $\{1,\ldots,n\}^{\zzz}$ to $\MM$. In general the structure
of the subset $\Sigma$ is very complex but for Anosov systems it is
possible to construct particular partitions for which the set $\Sigma$ can
be described easily. Given a partition $\QQ$ we call {\it
compatibility matrix} the $n\times n$ matrix $C$ given by $C_{ij}=1$
if ${\rm int}(S\QQ_i)\cap{\rm int}(\QQ_j)
\not= \emptyset$ and 0 otherwise. We say that
$\QQ$ is a {\it Markov partition} if the set $\Sigma$ is formed by the 
sequences
$\s$ such that $C_{\s_i,\s_{i+1}}=1$ for every $i\in {\ZZZ}$. This
means that the sequences that satisfy the above nearest neighbor
condition code all the points of $\MM$. In such a case we will denote
$\Sigma=\{1,\ldots,n\}^{\zzz}_C$.  

We now show how to construct a Markov partition for our model. We
start with $s_0$. A Markov partition $Q=\{Q_i, i=1,\ldots, n\}$ for
$s_0$ acting on $\TTT^2$ can be easily constructed starting from its
stable and unstable manifolds. Such a construction is standard and can
be found, \eg, in \cite{GBG03}. Let $C$ be its compatibility matrix
and $\hat c_0$ the associated symbolic dynamics.

It is important to note that $\hat c_0$ is Hold\"er continuous in the
sense that there exist constants $c$ and $\beta$ such that, for any
two sequences $\s,\s'\in\{1,\ldots,n\}^{\zzz}_C$, $d(\hat c_0(\s), \hat
c_0(\s'))\leq c \tilde d(\s,\s')^\beta$, with $\tilde
d(\s,\s')=e^{-\#(\s,\s')}$ where $\#(\s,\s')$ the biggest
integer such that $\s_j=\s_j'$, $\forall |j|\le \#(\s,\s')$. In
this case we can take $\beta=\ln(\lambda_+)$. Another key property is
that $C$ is a mixing matrix, this means that there exists a {\it
decorrelation time} $a\in \NNN$ such that $C^{a}$ has all entries strictly
positive. This means that we can connect any two element of the Markov
partition in $a$ time steps.

For every point $ s=\{s_\x\}_{\x\in \zzz^d}\in\{1,\ldots,n\}^{\zzz^d}$
we can consider the Cartesian product $\QQ_{ s}=\times_{\x\in
\zzz^d}Q_{s_\xi}\subset\TT_N$. It is clear that the collection of $\QQ_{
s}$ with $ s \in \{1,\ldots,n\}^{\zzz^d}$ forms a Markov partition for
$S_0$. Note that it is natural to index the element of this partition
with the element of $\{1,\ldots,n\}^{\zzz^d}$ so that we can associate to
this partition the symbolic dynamics $c_0:\ZZZ^d\times
\ZZZ=\ZZZ^{d+1}\to \TT_N$ naturally defined from $\hat c_0$. We
can still call $C$ the compatibility matrix and
$\{1,\ldots,n\}^{\zzz^{d+1}}_C$ the set of possible sequences
(namely $\s=\{\s_{\x,i}\}_{\x\in \zzz^d, \,i\in \zzz}$ is in
$\{1,\ldots,n\}^{\zzz^{d+1}}_C$ if and only if
$C_{\s_{\x,i}\s_{\x,i+1}}=1$ for every $\x\in \zzz^d$ and $i\in \ZZZ$).
Given any point $(\x,i)\in\ZZZ^{d+1}$ we will call $\x$ its {\it space
component} and $i$ its {\it time component}.

The key observation is now that the sets $h_\e(\QQ_{s})$ form a Markov
partition for $S_\e$. This implies that the space of symbolic
sequences for $S_\e$ is the same as that for $S_0$ and that the
symbolic dynamics $c_\e$ for $S_\e$ is given by
$c_\e(\s)=h_\e(c_0(\s))$. Clearly $c_\e$ is still H\"older
continuous. This completes the construction of the Markov partition
for $S_\e$. We thus obtained that the manifold $\TT$ can be mapped to
$\{1,\ldots,n\}^{\zzz^{d+1}}_C$ where $d$ directions of the lattice
$\ZZZ^{d+1}$ represent the $d$ directions of $\TT=(\TTT^2)^{\zzz^d}$
and the last represents the time evolution. Indeed the map $S_\e$ on
the space $\{1,\ldots,n\}^{\zzz^{d+1}}_C$ becomes the shift on the
time direction, to be called $\t$.

\subsection{SRB measure}\lb{ss2.3}

Let now consider the SRB measure $\m_N^{SRB}$ as defined in section
1. In this case we need to keep $N$ finite because it is not easy to
give a meaning or construct directly the SRB measure for $N=\infty$.

Let $m^{\SRB}_N$ be the measure on
$\{1,\ldots,n\}^{V_N\times\zzz}_C$ defined as
$m^{\SRB}_N(A)=\mu^{\SRB}_N(c_\e^{-1}(A))$, \ie $m^{\SRB}_N$ is the
image of $\m^{\SRB}_N$ via symbolic dynamics $c_\e$. The measure
$m^{\SRB}_N$ can be described efficiently through its restrictions to
finite subsets of $V_N\times\ZZZ$.

Given $\L\subset V_N\times\ZZZ$, $m^{\SRB}_N(\s_\L|\s_{\L^c})$ will
denote the probability of the event $\{\s'|\s'_\L=\s_\L\}$ 
conditional to the event
$\{\s'|\s'_{\L^c}=\s_{\L^c}\}$ w.r.t. to the probability measure 
$m^{\SRB}_N$, where
$\L^c=(V_N\times\ZZZ)\setminus \L$ and $\s_\L$ is the collection of
the $\s_{\x,i}$ for $(\x,i)\in \L$. 

>From the theory of SRB measures, see \cite{Si},
\cite{GBG03}, it follows that $m^{\SRB}_N$ is a Gibbs measure and its
conditional probabilities satisfy:
\be\frac
 {m^{\SRB}_N(\s'_\L|\s_{\L^c})}
 {m^{\SRB}_N(\s''_\L|\s_{\L^c})}
 = \lim_{K\to\io}\left[\frac
 {\DD_\e^{u(2K)}\big(c_\e(\t^{-K}\s')\big)}
 {\DD_\e^{u(2K)}\big(c_\e(\t^{-K}\s'')\big)} \right]^{-1}\;,\lb{2.3}\ee
where $\s'$ (resp. $\s''$) is the configuration coinciding with
$\s_\L'$ (resp. $\s_\L''$) on $\L$ and with $\s_{\L^c}$ on $\L^c$;
$\t$ is the image of $S_\e$ through $c_\e$ (\ie it is the one step shift
in time direction);
$\DD_\e^{u(n)}(\psi)$ measures the expansion of the volume on the
unstable manifold at the point $\psi$. To be more precise let
$\{w_{\e,+}^{(\x)}(\psi)\}_{\x\in V_N}$ be a basis on
$E^+_{\e,\psi}$. We will construct one such a basis in section 4. Then we
have 
\be \DD_\e^{u(n)}(\psi)\defin\sqrt{\frac
 {\det_{\x\h}\big[(DS_\e^n\ w_{\e,+}^{(\x)})\cdot
  (DS_\e^n\ w_{\e,+}^{(\h)})\big]}
 {\det_{\x\h}\big[w_{\e,+}^{(\x)}\cdot
  w_{\e,+}^{(\h)}\big]}}(\psi)\;,\lb{2.4}\ee
where $u\cdot v$ represent the usual scalar product in $\RRR^{V_N}$ and
$\det_{\x\h}$ is the determinant of the expression in square brackets
thought as a matrix indexed by $\x$ and $\h$.

Using the invariance of $E^+_\e$ under $S_\e$ and introducing
the {\it unstable Lyapunov matrix} $\LL(\psi)$ satisfying the
equation
$$DS_\e(\psi)w^{(\x)}_{\e,+}(\psi)=
\sum_\h w^{(\h)}_{\e,+}\big(S_\e(\psi)\big)\LL^{\h\x}(\psi),$$
we can rewrite the above
expression as 
\be
\DD_\e^{u(2K)}(S_\e^{-K}(\psi))=\frac
{\sqrt{\det_{\x\h}
\big(w^{(\x)}_{\e,+}\cdot w^{(\h)}_{\e,+}
\big)}(S_\e^K(\psi))}
{\sqrt{\det_{\x\h}\big(w^{(\x)}_{\e,+}\cdot
w^{(\h)}_{\e,+}\big)}(S_\e^{-K}(\psi))}
 \prod_{j=-K}^{K-1}\Big|
 \mbox{det}_{\x\h}
 \big[\LL^{\x\h}(S_\e^j\psi) \big]\Big|\;.\lb{2.5}\ee
Now the first ratio in eq.\pref{2.5}, when inserted in eq.\pref{2.4},
is vanishing ; indeed the uniform H\"older continuity of
$w^{(\x)}_{\e,+}(h_\e(\psi))$ and the fact that $\s'$ and $\s''$
are asymptotically identical in the past and in the future implies
that: 
\be \lim_{K\rightarrow\pm\infty}
 \Big(\ln\sqrt{\det\!_{\x\h}\big(w^{(\x)}_{\e,+}
 \cdot w^{(\h)}_{\e,+}\big)}
 \Big)\big(c_\e(\tau^K\s')\big)-
 \Big(\ln\sqrt{\det\!_{\x\h}\big(w^{(\x)}_{\e,+}
 \cdot w^{(\h)}_{\e,+}\big)}
 \Big)\big(c_\e(\t^K\s'')\big)=0\; ;\lb{2.7}\ee
thus the choice of the basis in $E^+$ does not change the
result, namely the SRB measure does not depend on the choice of the
metric as is to be expected from its definition.
Calling
$\L^\x(\psi)\defin\Big(\ln \LL(h_\e(\psi))\Big)^{\x\x}$ 
we finally get:
\be \frac
 {m^{\SRB}_N(\s'_\L|\s_{\L^c})}
 {m^{\SRB}_N(\s''_\L|\s_{\L^c})}=\exp
 \Bigg\{-\sum_{j=-\io}^{+\io}\sum_{\x\in V_N}
 \Bigg[\L^\x\big(c_0(\t^j\s')\big)-
 \L^\x\big(c_0(\t^j\s'')\big) \Bigg]
 \Bigg\}\;.\lb{2.6}\ee
Here we used the the fact that $c_\e=h_\e\circ c_0$. 
Furthermore the H\"older continuity of $\L^\x\big(c_0(\s')\big)$ implies
absolute convergence of the sum in eq. \pref{2.6} because only points
asymptotically equal both in the past and in the future are compared.

The crucial point of this construction is that
the matrix $L(\psi)\defin\LL\big(h_\e(\psi)\big)$ 
is analytic in $\e$ due to the fact that
depends only on $w^{(\x)}_{\e,+}(h_\e(\psi))$. As we already
noted $w^{(\x)}_{\e,+}(h_\e(\psi))$ are analytic in $\e$. We
will prove this fact in section 4.

In section 6 we will apply to eq. \pref{2.6} the standard methods
developed in the study of Gibbs measure in Statistical Mechanics. To
do this we will need to decompose the ``interaction''
$\L^\x\big(c_0(\s)\big)$ as the sum of potentials depending only on
$\s_X\defin\{\s_j\}_{j\in X}$ 
where $X$ is a finite subset of $\ZZZ^{d+1}$. More
precisely we will decompose:
\be 
 \sum_{(\x,i)\in V_N\times\zzz}
 \L^\x\big(c_0(\t^i\s)\big)
 =
 \sum_{X\subset V_N\times\zzz}
 \phi_X(\sigma_X);\lb{2.11}\ee
(these two series are not convergent: they represent
the  formal expression for the ``Hamiltonian'' of 
a Gibbs measure. See section \pref{ssL} for a more precise statement). 
We shall show that we can choose
$\phi_X$ analytic in $\e$,
translationally invariant in space and time directions
and decaying exponentially in the {\it tree distance} of the set $X$, namely 
the length of the shortest tree connecting all the lattice points in $X$. 
In this way  \pref{2.6} can be written as
\be \frac
 {m^{\SRB}_N(\s'_\L|\s_{\L^c})}
 {m^{\SRB}_N(\s''_\L|\s_{\L^c})}=\exp\Bigl\{-\sum_{X\cap 
\L\neq\emptyset}\left[\phi_X(\sigma_X')-\phi_X(\s_X'')\right]\Bigr\}
\;,\lb{2.8a}\ee
so that one can finally write:
\be 
 m^{\SRB}_N(\s_\L|\s_{\L^c})=
 \frac{\exp\big\{-\sum_{X\cap \Lambda\neq\emptyset}
\phi_X(\sigma_X)\big\}}
{\sum_{\sigma_\Lambda}\exp\big\{-\sum_{X\cap 
\Lambda\neq\emptyset}\phi_X(\sigma_X)\big\}}\;.\lb{2.9}\ee
This will allow us to show our
analyticity claim uniformly in $N$. 

\section{Perturbative Construction of the SRB measure}\lb{s3}\setcounter{equation}{0}
   
In this section we construct the conjugation $h_\e$ and prove that it
is analytic in $\e$. The technique we use consists in expanding $h_\e$
as a power series in $\e$ and writing a
recursive relation linking the $n$--th order coefficient to the 
coefficients of order
$i$ with $i<n$. This naturally leads to a tree expansion of the form
usual in perturbation theory for Quantum Field Theory, \ie the tree we
will introduce are the ``Feynmann graphs'' of our theory. See also \cite{GBG03}
and reference therein for similar application to KAM theory. 

\subsection{The Conjugation}\lb{ss3.1}

>From now on we will identify functions from $\TT_N$ to $\TT_N$ with their
lifts as functions from $\RRR^{2V_N}$ to $\RRR^{2V_N}$. 
Using the definition \pref{2.1} and looking for
$h_\epsilon$ of the form $h_\epsilon(\psi)=\psi+\delta
h_\epsilon(\psi)$, we find:
\be
\delta h_\e\circ S_0-S_0\circ \delta h_\e=
 \e f\circ({\rm Id}+\delta h_\e)\;.\lb{3.1}
\ee
where ${\rm Id}$ is the identity map. 

Setting $\lambda\defin\lambda_-=\lambda_+^{-1}$ and writing
$f(\psi)\defin \sum_{\xi,\a} f^{\x^\a}(\psi)w_{0,\a}^{(\x)}$ and 
similarly for $\delta
h_\e^{\x^\pm}$, we get:
\be \left\{ \begin{array}{rl}
   \delta h_\e^{\x^+}(S_0\psi)
   -
   \lambda^{-1} \delta h_\e^{\x^+}(\psi)
  &=
   \e f^{\x^+}\Big(\psi + \delta h_\e(\psi)\Big)
   \\ 
   \\
   \delta h_\e^{\x^-}(S_0\psi)
   -
   \lambda \delta h_\e^{\x^-}(\psi)
  &=
   \e f^{\x^-}\Big(\psi + \delta h_\e(\psi)\Big)
   \end{array}\right.\; .\lb{3.2}\ee
Both equations can be implicitly solved by iteration:
\be \delta h_\e^{\x^\a}(\psi)
 =
 -\a\e \sum_{p\geq0} \lambda^{p+\rho_\a}
 f^{\x^\a}\Big(S_0^{\a(p+1-\rho_\a)}\psi + 
 \delta h_\e(S_0^{\a(p+1-\rho_\a)}\psi)\Big)\;,\lb{3.3}\ee
where $\rho_\a=(1+\a)/2$.

It is easy to see that the series in eq.\pref{3.3} is absolutely
convergent, since $\lambda<1$ and $f$ is bounded. 
Expanding $f^{\x^\a}\big(\psi + \delta h_\e(\psi)\big)$
in power of its argument we find:
\be f^{\xx}\big(\psi + \delta h_\e(\psi)\big)
 =f^{\xx}(\psi)+
 \sum_{k\geq1}\e^k\sum_{s=1}^k
 \sum_{k_1,\dots,k_s \atop k_j\geq1}^{k_1+\dots+ k_s=k}
 \Big(
  \frac{f^{\xx\!,\,\xx_1\dots\,\xx_s}}
       {s!}
  \delta h_{(k_1)}^{\xx_1}\cdots \delta h_{(k_s)}^{\xx_s}
 \Big)(\psi)\;,\lb{3.4}\ee
where we have introduced the index $\xx=(\xi,\alpha)$, with
$\alpha=\pm$, and
$f^{\xx\!,\,\xx_1\dots\,\xx_s}=\partial_{\xx_1}\cdots\partial_{\xx_s}
f^{\xx}$ with $\partial_{(\x^\a)}$ the partial derivative in the
direction of $w_{0,\alpha}^{(\xi)}$. Moreover we use the convention of
summing on twice repeated indexes. The first order coefficient of the
expansion of the conjugation is then:
\be \delta h^{\xx}_{(1)}(\psi)=
  (-\a)
 \sum_{p\geq0}
 \lambda^{p+\rho_\a} f^{\xx}
 (S_0^{\a(p+1-\rho_\a)}\psi)\;,\lb{3.5}\ee
while the $k$--th, $k> 1$, coefficient turns out to be:
\be \delta h^{\xx}_{(k)}(\psi)=
 \sum_{s=1}^k
 \sum_{k_1,\dots,k_s \atop k_j\geq1}^{k_1+\dots k_s=k-1}
 (-\a)
 \sum_{p\geq0}
 \lambda^{p+\rho_\a}
 \Big(
  \frac{f^{\xx,\,\xx_1\dots\,\xx_s}}
       {s!}
  \delta h_{(k_1)}^{\xx_1}\cdots \delta h_{(k_s)}^{\xx_s}
 \Big)(S_0^{\a(p+1-\rho_\a)}\psi)\;.\lb{3.6}\ee
>From eq.\pref{3.5} we see that $\delta h_\e$ is in general non
differentiable with respect to $\psi$. Indeed already differentiating
$\d h^{\xi^+}_{(1)}(\psi)$ with respect to $\psi$ we find a non
converging series. On the contrary it is clear that  eq.\pref{3.5} is
H\"older continuous in $\psi$ for every exponent $\beta<1$.

We can interpret eq.\pref{3.6} graphically as shown in Fig. \ref{f1}.

\insertplot{300}{100}{ \ins{-90pt}{42pt}{$\xx$}
\ins{-101pt}{25pt}{$r$} \ins{-73pt}{23pt}{$v$}
\ins{-65pt}{35pt}{$(k)$} \ins{-35pt}{33pt}{$=$}
\ins{-15pt}{39pt}{$\sum_{s_v\ge 1}^{k_1+\cdots k_s=k-1}$}
\ins{67pt}{27pt}{$r$} \ins{80pt}{42pt}{$\xx$} \ins{95pt}{25pt}{$v$}
\ins{103pt}{62pt}{$\xx_1$} \ins{99pt}{10pt}{$\xx_{s_v}$}
\ins{139pt}{78pt}{$(k_1)$} \ins{130pt}{65pt}{$v_1$}
\ins{139pt}{58pt}{$(k_2)$} \ins{130pt}{45pt}{$v_2$}
\ins{130pt}{40pt}{$\vdots$} \ins{139pt}{18pt}{$(k_{s_v-1})$}
\ins{130pt}{5pt}{$v_{s_v-1}$} \ins{139pt}{-2pt}{$(k_{s_v})$}
\ins{130pt}{-15pt}{$v_{s_v}$} } {fig1} {Graphical interpretation of
\pref{3.6}. \lb{f1}}{0}

The l.h.s. of the graphical equation in Figure \ref{f1} represents $\d
h^{\xx}_{(k)}(\psi)$ while the r.h.s., representing the sum in eq.
\pref{3.6}, is a ``simple tree'' consisting of a ``root'' $r$,
a ``root branch'' $\l_v\=(r,v)$ coming from the ``node'' (or ``vertex'')
$v$ and $s_v$ branches ``entering $v$'', to be called $\l_{v_i}\=(v,v_i)$, 
$i=1,\ldots,s_v$.

Even if the drawing in the figure does not carry them explicitly, we
imagine that some labels are affixed to the node $v$: more precisely 
$\xx(v)=(\x(v),\a(v))\in V_N\times\{\pm\}$ and $p_v\in\ZZZ_+$. 
Furthermore a label $\xx_\l=(\x_\l,\a_\l)
\in V_N\times\{\pm\}$ is associated to each branch $\l$. In the 
figure above $\xx_{\l_v}\=\xx$ and $\xx_{\l_{v_i}}\=\xx_i$, 
$i=1,\ldots,s$.

The node $v$ symbolizes the tensor with entries 
\be N_{v;\xx,\xx_1,\ldots,\xx_s}\defin (-\a(v))\l^{p_v+\r_{\a(v)}}{f^{\xx,
\xx_1\ldots\xx_{s_v}}\over s_v!}(S_0^{p(v)}\psi)\;,\lb{3.6a}
\ee
where $p(v)=\a(v)(p_v+1-\r_{\a(v)})$. Observe that, in order for eq. 
\pref{3.6a} and \pref{3.6} to be non zero, we
must have $|\xi_{\l_{v_i}}-\xi(v)|\leq 1$, 
due to our definition of the coupling $f$.

The line $\l_v$ exiting vertex $v$ symbolizes the {\it propagator}, that
is simply $\d_{\xx_{\l_v},\xx(v)}$.

The line with label $\xx$ exiting from the bullet of the l.h.s. with
label $(k)$ represents 
$\d h^\xx_{(k)}(\psi)$; the branches with labels $\xx_i$ 
exiting from the bullets of the r.h.s. with label $(k_i)$ represent 
$\d h^{\xx_i}_{(k_i)}(S_0^{p(v)}\psi)$.

Even if it is not explicitly written in the figure above, a summation over
the free indices $\xx(v),\xx_{\l_{v_i}}$ has to be performed (note that
the summation over $\xx(v)$ simply fixes $\xx(v)=\xx$, because of
the presence of the propagator $\d_{\xx(v),\xx}$).

Since eq. \pref{3.6} is multilinear in $\d h^{\xx_i}_{k_i}$,
we can just replace each of the branches exiting from a bullet with the 
same graphical expression in the r.h.s. of the above figure. And so
on, until the labels $(k_i)$ on the bullets 
({\it top nodes}) become equal 
to $1$. In this case the end--points represent $\d h_{(1)}$, that is a 
known expression, see eq. \pref{3.5}, and we will draw these known end--points
as small dots. 

Thus we have represented our $\d h^\xx_{(k)}$ as a ``sum over trees'' 
with $k$ branches and $k$ nodes (we shall not regard the root as a node)
of suitable {\it tree values}. In Fig. \ref{f2} we draw a typical tree $\th$
we get via such procedure.

\insertplot{300}{150}{ \ins{-83pt}{65pt}{$r$} \ins{-13pt}{65pt}{$v_0$}
\ins{67pt}{115pt}{$v_1$} \ins{67pt}{26pt}{$v_2$}
\ins{107pt}{80pt}{$v_3$} \ins{155pt}{143pt}{$v_4$}
\ins{155pt}{103pt}{$v_5$} \ins{155pt}{63pt}{$v_6$}
\ins{155pt}{43pt}{$v_7$} \ins{155pt}{23pt}{$v_8$}
\ins{155pt}{-7pt}{$v_9$} } 
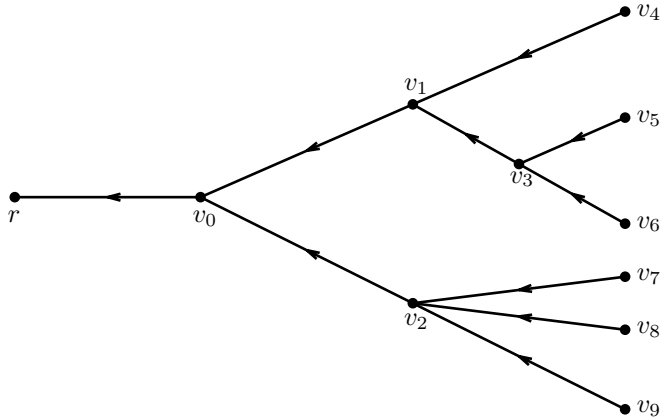
{fig7} {A tree $\th$ of order $k=10$
appearing in the expansion for $\d h_\e$. Labels $\x(v_i)$, $\a(v_i)$
and $p_{v_i}$ are associated to all vertices $v_i$. \lb{f2}}{0}

Note that a tree induces a partial ordering among
its nodes: a node $w$ precedes $v$ (and it will be written $w<v$) if
there is a path of branches connecting $w$ and $v$ with the arrows
pointing from $w$ to $v$. 

Let us now summarize the discussion above. Let 
$\widehat\TT_k(\xx)$ be the set of rooted trees with
$k$ branches and $k$ nodes, with labels $\xx(v),p_v$ attached to their
vertices and $\xx(v_0)\=\xx$, where $v_0$ is the last vertex 
preceding the root. Given $\th\in \widehat\TT_k(\xx)$, let
the value of $\th$ be defined as
\be \widehat\Val(\th,\psi)=\prod_{v\in\th}
 \Big(-\a(v)\Big)
 \l^{p_v+\rho_{\a(v)}}
  \frac{f^{\xx(v),\;\xx(v_1)\dots\;
  \xx(v_{s_v})}}{s_v!}
 (S_0^{p(v)}\psi)\;,\lb{3.7}\ee
where $v_1,\ldots,v_{s_v}$
are the nodes immediately preceding $v$ and $p(v)=\sum_{w\ge v}
\a(v)(p_v+1-\r_{\a(v)})$. With these definitions $\d h^{\xx}_{(k)}
(\psi)$ can be calculated as $\d h^{\xx}_{(k)}
(\psi)=\sum_{\th\in\widehat\TT_k(\xx)}\widehat\Val(\th,\psi)$.

\subsection{Convergence 
and Regularity of the Perturbative Expansion of the Conjugation}

By definition $g(\psi)$ depends only on $\psi_{nn(0)}$ so that 
it is analytic in $\DD\defin\{\psi^i_\x\in\CCC|\ |\Im\psi^i_\x|\le
r_0,\ i=1,2,\ \x\in nn(0)\}$ for some $r_0>0$.
Calling $G$ the maximum of $g$ on $\DD$, from
Cauchy's formula we get
\be  |f^{\xx,\;\xx_1\dots
  \xx_s}(\psi)|
  \le G \frac{m_1!\cdots m_D!}{r_0^s}
  \le G \frac{s!}{r_0^s}.
\lb{3.7a}\ee
where $m_1,...m_D$ are the multiplicities  of the partial derivatives
with respect to the $D\defin 2(2d+1)=2|nn(0)|$ possible variables, (thus
$m_1+...+m_D=s$).
 
In the same way, if $\psi$ and $\psi'$ are
identical on each site but $\x'\in nn(\x)$ and if $0<\b\le 1$ we get
%
\be  |f^{\xx,\xx_1\dots\
  \xx_s}(S_0^{p}\psi)-f^{\xx,\xx_1\dots
  \xx_s}(S_0^{p}\psi')|\le 
   G \frac{(s+1)!}{r_0^{s+1}}
   (2\p^2)^{\frac{1-\b}{2}}\l^{-\b p}
  |\psi_{\x'}-\psi'_{\x'}|^\b\ee
where we have used the periodicity of $f$.  Next we bound the value of
a tree $\th\in\widehat\TT_k(\xx)$. Using eq. \pref{3.7a}, for $\th\in
\widehat\TT_k(\xx)$, we find:
\be ||\widehat\Val(\th,\cdot)||_\io
\le\prod_{v\in\th}\l^{p_v}{G\over r_0^{s_v}}=
{G^k\over r_0^{k-1}}\prod_{v\in\th}\l^{p_v}\;,\lb{val}\ee
where we used that, if $\th\in\widehat\TT_k(\xx)$, $\sum_{v\in\th}s_v=k-1$.

The sum over the trees can be interpreted
as a sum over the topological trees and a sum over the labels attached to the
trees. If $\Theta_k$ is the set of topological trees of order $k$, we get:
\bea &&||\d h_{(k)}^{\xx}||_\io\le \sum_{\th\in\Theta_k}\sum_{\xx(v),\a(v)\atop
v\in\th}{G^k\over r_0^{k-1}}\sum_{p_v\atop v\in\th}\prod_{v\in\th}\l^{p_v}
= \sum_{\th\in\Theta_k}\sum_{\x(v),\a(v)\atop
v\in\th}{G^k\over r_0^{k-1}}({1\over 1-\l})^k\le\lb{he}\\
&&\le\sum_{\th\in\Theta_k}2^k(2d+1)^k{G^k\over r_0^{k-1}}({1\over 1-\l})^k\le
2^{2k}2^k(2d+1)^k{G^k\over r_0^{k-1}}({1\over 1-\l})^k\;,\nn\eea
where we used that:
\begin{enumerate}

\item{$2^k$ is the number of terms in the sum over 
the $\a(v)$ indices;}

\item{$(2d+1)^k$ is a bound on the number 
of terms in the sum over the values of the $\x(v)$ indices
not making $\widehat\Val(\th,\psi)$ vanish (observe that, given a 
tree $\th$, its value $\widehat\Val(\th,\psi)$ 
is vanishing unless $|\x(v')-\x(v)|\le 1$, where $v'$ is the node 
immediately preceding $v$);}

\item{$2^{2k}$ is a bound on the number of unlabeled 
rooted trees with $k$ nodes.}

\end{enumerate} 

In the same way we find that, if $\psi$ and $\psi'$ are identical on 
each site but $\x'$ and if $0<\b<1$:
\bea 
{|\d h_{(k)}^{\xx}(\psi)-
\d h_{(k)}^{\xx}(\psi')|\over |\psi_{\x'}-\psi_{\x'}'|^\b}
&&\le \sum_{\th\in\Theta}\left({1\over 1-
\l^{1-\b}}\right)^{k}(2\p^2)^{{1-\b\over 2}}\frac{G^k}{r_0^k}2^k
(2d+1)^k\sum_{v\in\th}(s_v+1)\le\lb{3.8aa}\\
&&\le 2^{2k}\left({1\over 1-
\l^{1-\b}}\right)^{k}(2\p^2)^{{1-\b\over 2}}\frac{G^k}{r_0^k}2^k
(2d+1)^k(2k-1)\;.\nn\eea
So the map $h_\e:\TT_N\to\TT_N$ exists, it is h\"older continuous 
w.r.t. $\psi$ and analytic w.r.t. $\e$ in the complex 
disc $|\e|\le\e_\b$, with 
\be\e_\b=\left[{1\over 1-
\l^{1-\b}}2^3\frac{G}{r_0}
(2d+1)\right]^{-1}\;.\lb{3.8a}\ee
In order to prove that $h_\e(\psi)$ is an {\it homeomorphism}, we have
to show that it is invertible. The proof is easy and standard.
Regarding injectivity, note that two {\it distinct} points $\psi_1$,
$\psi_2$, are necessarily far order one in the ``future'' or in the
``past'', namely there exists an integer $n\in\ZZZ$ such that
$|S_0^n\psi_1-S_0^n\psi_2|=O(1)$; then
$S_\e^n\Big(h_\e(\psi_1)\Big)-S_\e^n\Big(h_\e(\psi_2)\Big)
=S_0^n(\psi_1-\psi_2)+\d h_\e(S_0^n\psi_1)-\d h_\e(S_0^n\psi_2)$,
cannot vanish as the first term is order one, the other two of order
$\e$; thus it cannot be but $h_\e(\psi_1)\neq h_\e(\psi_2)$.
Regarding surjectivity, since $f$ is a continuous injective mapping on
a torus, $f$ is necessarily surjective (the proof is trivial on
$\TTT^1$ and it can be easily extended by induction to $\TT_N$).
 
\section{The Unstable Direction}\lb{ss3.2}\setcounter{equation}{0}

In order to explicitly compute the SRB measure, we have to construct a
basis for the unstable subspace $E^+_\psi$, and the expansion
coefficient $\DD_\e^{(n)}$ associated to it, as explained in
section \ref{ss2.2} above.  Note that we cannot use $h_\e$ to find a basis
for $E^+_\psi$ because it is only H\"older continuous.

To find the unstable base $\{{w}_{\e,+}^{(\x)}(\psi)\}_{\x\in V}$ and
the {\it Lyapunov matrix} $\mathcal{L}(\psi)$ we have to solve the
following equation
\be\Big(DS_\e \ {w}_{\e,+}^{(\h)}\Big)(\psi)=
 {w}_{\e,+}^{(\x)}\Big(S_\e(\psi)\Big)
 \mathcal{L}^{\x\h}(\psi)\;.\lb{3.9}\ee
In general this equation cannot have solutions analytic in $\e$. In
fact, from the general theory we know that the unstable vectors
$\{w_{\e,+}^{(\x)}(\psi)\}_{\x\in V}$ are not differentiable
with respect to $\psi$.  But, as we previously pointed out, to compute
the SRB measure we need only to know the expansion coefficient at the
point $h_\e(\psi)$, \ie $\DD_\e^{(n)}(h_\e(\psi))$. Let us define
$w_{\e,+}^{(\x)}(h_\e(\psi))
\defin v^{(\x)}_\e(\psi)$ for $\x\in V$ and note that
${v}^{(\x)}_\e(\psi)$ satisfies the equation
\be\big(DS_\e\big)\big(h_\e(\psi)\big)v_\e^{(\h)}(\psi)=
 {v}_\e^{(\x)}\left(S_0\psi\right)
 L^{\x\h}(\psi)\virg L(\psi)=\LL(h_\e(\psi))\;.\lb{3.10}\ee
We will show that this equation admit a solution analytic in
$\e$. Moreover the determinant of $L(\psi)$ is all what we need to
compute the SRB measure.

At this point, it is convenient to write eq.\pref{3.10} in components. 
Denoting by $\yy$ the double index $\h^\b$ (again
$\xx\defin\x^\a$), defining 
$v^{(\x)}_\e(\psi)\defin \sum_\h V_{\e,\yy}^{(\x)}(\psi)
w_{0,\b}^{(\h)}$ and  
$\big(DS_\e\ w_{0,\b}^{(\h)}\big)(\psi)\defin \sum_\xx S_\e^{\xx,\yy}(\psi)
w_{0,\a}^{(\x)}$, we get
\be S_\e^{\xx,\yy}\Big(h_\e(\psi)\Big) V_{\e,\yy}^{(\r)}(\psi)=
 V_{\e,\xx}^{(\z)}(S_0\psi)L^{\z\r}(\psi)\;.\lb{3.10a}\ee
Now, defining the corrections $\d L$ and $\d V$ as follows:
\be L^{\x\h}(\psi)\defin\l^{-1}\d_{\x\h}+\d L^{\x\h}(\psi)\virg
V_{\e,\xx}^{(\x)}(\psi)\defin V_{0,\xx}^{(\x)}+\d V_{\xx}^{(\x)}(\psi)
\quad{\rm with}\quad V_{0,\h^+}^{(\x)}=\d_{\x,\h}\virg 
V_{0,\h^-}^{(\x)}=0\;,\lb{3.11}\ee
we find that \pref{3.10a} is equivalent to:
\be \left\{ \begin{array}{rl}
 &\delta L^{\x\r}(\psi)
   = 
  \l^{-1}
  \big[\d V^{(\r)}_{\x^+}(\psi)-\d V^{(\r)}_{\x^+}(S_0\psi)\big]+
   \e f^{\x^+,\r^+}(h_\e(\psi))+\nn\\
 &\phantom{\delta L_{\x\r}(\psi)= }
  +\e f^{\x^+,\yy}(h_\e(\psi))\d V_\yy^{(\r)}(\psi)
  -\d V_{\x^+}^{(\z)}(S_0\psi)\d L^{\z\r}(\psi)\lb{3.12}\\
  \\
 &\l\d V_{\x^-}^{(\r)}(\psi)-\l^{-1}\d V_{\x^-}^{(\r)}(S_0\psi)=
  -\e f^{\x^-,\r^+}(h_\e(\psi))-\e f^{\x^-,\yy}(h_\e(\psi))\d V_{\yy}^{(\r)}
(\psi)+\nn\\
 &\phantom{\l\d V_{\x^-}^{(\r)}(\psi)-\l^{-1}\d V_{\x^-}^{(\r)}(S_0\psi)=}
 +\d V_{\x^-}^{(\z)}(S_0\psi)\d L^{\z\r}(\psi)\;.\end{array}
  \right.\nn\ee 
Of course the above equations cannot determine completely the basis and
its associated matrix: indeed given a solution $\{V_\yy^{(\r)}(\psi)\}$,
$\{L^{\z\r}(\psi)\}$ of eq.\pref{3.10} and a generic invertible H\"older
continuous matrix $R^{\g\r}(\psi)$, also $\{V_\yy^{(\g)}(\psi)R^{\g
\r}(\psi)\}$, $\{R^{-1,\z\d}(S_0\psi)L^{\d\g}(\psi)R^{\g\r}(\psi)\}$
solve \pref{3.10}. Thus it is possible to add a constrain to $\d
V_{\yy}^{(\r)}(\psi)$: a possible choice, which simplifies a lot the
expressions above, consists in taking $\d V_{\r^+}^{(\x)}(\psi)=0$, so that
\pref{3.12} becomes:
\be \left\{ \begin{array}{rl}
 &\delta L^{\x\r}(\psi)=\e f^{\x^+,\r^+}(h_\e(\psi))+\e f^{\x^+,\h^-}
  (h_\e(\psi))\d V_{\h^-}^{(\r)}(\psi)\lb{3.13}\\
  \\
 &\l\d V_{\x^-}^{(\r)}(\psi)-\l^{-1}\d V_{\x^-}^{(\r)}(S_0\psi)=
  -\e f^{\x^-,\r^+}(h_\e(\psi))-\e f^{\x^-,\h^-}(h_\e(\psi))\d V_{\h^-}^{(\r)}
  (\psi)+\nn\\
 &\phantom{\l\d V_{\x^-}^{(\r)}(\psi)-\l^{-1}\d V_{\x^-}^{(\r)}(S_0\psi)=}
  +\d V_{\x^-}^{(\z)}(S_0\psi)\d L^{\z\r}(\psi)\;.\end{array}
  \right.\nn\ee 
An implicit solution of \pref{3.13} (to be inverted iteratively by a new tree
expansion, see below) is:
\be \left\{ \begin{array}{rl}
  \delta L^{\x\r}(\psi)
 &=\e f^{\x^+,\r^+}(h_\e(\psi))+\e f^{\x^+,\h^-}
  (h_\e(\psi))\d V_{\h^-}^{(\r)}(\psi)\lb{3.14}\\
  \\
  \d V_{\x^-}^{(\r)}(\psi)
 &=\sum_{j\geq0}\l^{2j+1}
  \Big[\e f^{\x^-,\r^+}(h_\e(S_0^{-j}\psi))+\e f^{\x^-,\h^-}
  (h_\e(S_0^{-j}\psi))\d V_{\h^-}^{(\r)}
  (S_0^{-j}\psi)+\nn\\
 &\phantom{=\sum_{j\geq0}\l^{2j+1}\Big[}
-\d V_{\x^-}^{(\z)}(S_0^{-j+1}\psi)\d L^{\z\r}(S_0^{-j}
  \psi)\Big]\;\end{array}
  \right.\nn\ee
As for the construction of the conjugation, we can expand in
power series of $\e$ both sides of eq.\pref{3.14} and equate the
coefficients of the same order, thus finding an iterative solution
of $\d L_{(k)}$ and $\d V_{(k)}$. The first order coefficients are
given by:
\bea &&\delta L^{\x\r}_{(1)}(\psi)
     =f^{\x^+,\r^+}(\psi)\nn\\
     &&\d V_{\x^-(1)}^{(\r)}(\psi)
     =\sum_{j\geq0}\l^{2j+1}
  f^{\x^-,\r^+}(S_0^{-j}\psi)\;,\lb{3.15}\eea
while, for $k+1\ge 2$:
\bea 
 \d L_{(k+1)}^{\x\r}(\psi)=&&\sum_{s\ge 1,\ k_i\ge 1}^{
k_1+\cdots+k_s=k}
\Bigl(
{f^{\x^+,\r^+\xx_1\ldots\xx_s}\over s!}\d h^{\xx_1}_{(k_1)}
\cdots\d h^{\xx_s}_{(k_s)}\Bigr)(\psi)+\lb{3.15a}\\
+&&\sum_{s\ge 1,\ k_i\ge 1}^{
k_1+\cdots+k_s=k}
\Bigl({f^{\x^+,\h^-\xx_2\ldots\xx_{s}}\over (s-1)!}
\d V^{(\r)}_{\h^-(k_1)}\d h^{\xx_2}_{(k_2)}
\cdots\d h^{\xx_{s}}_{(k_{s})}
\Bigr)(\psi)\nn\eea
and
\bea\d V_{\x^-(k+1)}^{(\r)}(\psi)=
&&\sum_{s\ge 1,\ k_i\ge 1}^{
k_1+\cdots+k_s=k}\sum_{j\geq0}\l^{2j+1}
\Bigl({f^{\x^-,\r^+\xx_1\cdots \xx_s}\over s!}
  \d h_{(k_1)}^{\xx_1}\cdots \d h_{(k_s)}^{\xx_s}\Bigr)(S_0^{-j}\psi)+
\nn\\
 +&&\sum_{s\ge 1,\ k_i\ge 1}^{k_1+\cdots+k_s=k}
\sum_{j\geq0}\l^{2j+1}
\Bigl({f^{\x^-,\h^-\xx_2\cdots \xx_s}\over(s-1)!}\d V^{(\r)}_{\h^-(k_1)}
  \d h_{(k_2)}^{\xx_2}\cdots \d h_{(k_s)}^{\xx_s}
  \Bigr)(S_0^{-j}\psi)+\lb{3.16}\\
 -&&\sum_{s\ge 1,\ k_i\ge 1}^{k_1+\cdots+k_s=k}\sum_{j\geq0}\l^{2j+1}
 \Bigl({f^{\z^+,\r^+\xx_2\ldots\xx_s}\over (s-1)!}
(\d V_{\x^-(k_1)}^{(\z)}\circ S_0)\,
\d h^{\xx_2}_{(k_2)}
\cdots\d h^{\xx_s}_{(k_s)}\Bigr)(S_0^{-j}\psi)+\nn\\
 -&&\sum_{s\ge 2,\ k_i\ge 1}^{k_1+\cdots+k_s=k}\sum_{j\geq0}\l^{2j+1}
 \Bigl({f^{\z^+,\h^-\xx_3\ldots\xx_{s}}\over (s-2)!}
(\d V_{\x^-(k_1)}^{(\z)}\circ S_0)\,
\d V^{(\r)}_{\h^-(k_2)}\d h^{\xx_3}_{(k_3)}
\cdots\d h^{\xx_{s}}_{(k_{s})}\Bigr)(S_0^{-j}\psi)\;.\nn\eea
This two relations, together with \pref{3.6}, allow a recursive
construction of $\d L$ and $\d V$. Obviously, repeating the discussion
of \S\ref{ss3.1}, one finds that $\d L$ and $\d V$ can be expressed as
sums over trees, obtained by suitably modifying the construction of
previous section. It can be easily realized that the estimates for the
tree values are qualitatively the same as before, see
eq.\pref{val},\pref{he},\pref{3.8aa}. We point out the differences
appearing in the tree expansion for $\d V$:
\begin{enumerate}

\item the nodes can be of 4 different types (corresponding to the 4 lines
in eq. \pref{3.16}); so that the number of possible labels for a tree of
order $k$ is larger of a factor $4^k$;

\item the number $D_v$ of derivatives acting on a node function can be
either $s_v$ or $s_v+1$, see eq.\pref{3.15a},\pref{3.16},
so that $D_v!$ differs from the combinatorial factor $s_v!$ by at most
$s_v+1$; then the final estimate contains a factor that can be bounded by
$(1/r_0)\prod_v(s_v+1)\le e^k/ r_0$.
\end{enumerate}

A similar discussion can be made for the tree expansion of $\d L$.

The result is that $L$ and $V$ are analytic in $\e$ 
and H\"older continuous in $\psi$ with exponent $0<\b<1$ in a disc 
$|\e|\le\e_\b'$, with $\e_\b'$ smaller than the convergence radius $\e_\b$ 
of $h_\e$, see eq.\pref{3.8a}. Note that also in this case $\e_\b'$ is
independent of $N$.

As already explained (see section \ref{ss2.2} and in particular
eq.\pref{2.6}), in order to compute the SRB measure we need $\L^\x=
(\log L)^{\x\x}\defin -\log\l+\d\L^\x$, where:
\be \d\L^\x(\psi)=\Big[\log(
 I+\l\d L)\Big]^{\x\x}=\sum_{s\geq1}\frac{(-1)^{s+1}}{s}
 \l^s \d L^{\x\h_1}(\psi)\cdots\d L^{\h_{s-1}\x}(\psi)\;.\lb{3.17}\ee
(no summation on $\x$ is intended).
Expanding eq.\pref{3.17} in series of $\e$, we get:
\be \d\L^\x_{(k)}(\psi)=\sum_{s\ge 1,\ k_i\ge 1}^{k_1+\cdots k_s=k}
 \frac{(-1)^{s+1}}{s}
 \l^s\d L_{(k_1)}^{\x\h_1}(\psi)\cdots
 \d L_{(k_s)}^{\h_{s-1}\x}(\psi)\;.\lb{3.18}\ee
Again, the last equation, together with \pref{3.16} and \pref{3.6}, 
allows a recursive construction of the coefficients $\d\L^\x_{(k)}$
and the result is that $\L^\x$ is a sum over (suitably modified)
trees. The bounds are still qualitatively the same, so that
$\L^\x$ is analytic w.r.t. $\e$ in a suitably small complex disc
(independent of $N$) and H\"older continuous w.r.t. $\psi$.

   \section{SRB Potentials}\lb{ss3.3}\setcounter{equation}{0}

The next step towards the construction of the SRB measure and the proof
of its analyticity consists in the expansion of $\L^{\x}$ in potentials 
$\phi_X$. From the analysis of previous sections follows
that $\L^\x$, as well as  $h^\xx_\e$, $V^{(\r)}_{\x^-}$ and $L^{\x\h}$,
can be expanded in convergent sums over tree values. We will discuss here how
to expand $h$ in potentials, since the analogous expansion  
for $V$, $L$ and $\L$ is conceptually similar, just more
involved due to the more complex structure of the trees. 

We will proceed as follows. We first write the values of the trees in
terms of the symbolic variables $\s$. We then decompose each of these
values as a sum of terms  only depending on the $\s$'s on
finite but arbitrary large sets. Finally we define the associated
potentials by collecting together the contributions which depend on
the same $\s$'s.  Our goal is to obtain potentials defined over sets
with rather arbitrary shape but decaying exponentially with the {\it
tree distance} (see after eq.\pref{2.11} for a precise definition) 
of their support.

To begin with we expand the derivatives of the perturbation function
$f$ via a telescopic sum. Given the digits $s$ and $s'\in
\{1,\ldots,n\}$ we can always find a sequence of digits 
$\Sigma(s,s')=s_1s_2\ldots s_{a-1}$ such that the sequence
$s\Sigma(s,s')s'$
compatible, \ie such that $C_{s_i,s_{i+1}}=1$ for $i=0,\ldots
a-1$, where $s_0=s$ and $s_{a}=s'$. Chosen a sequence 
$\hat\s\in\{1,\ldots,n\}^{\zzz}_C$ once and for
all, given $\s\in \{1,\ldots,n\}^{\zzz}_C$ we can define its restriction
to time $j$, $\s^j$ as follows: $\s^j_{\x,t}=\s_{\x,t}$ if $|t|\leq
j$,  $\s^j_{\x,t}=\hat\s_{\x,t}$ if $|t|> j+a$ and the gap is
filled with the sequence constructed above for $s=\s_{\x,\pm j}$ and
$s'=\hat\s_{\x,\pm(j+a)}$. We can now define:
\bea f^{\xx,\xx_1\cdots\xx_s}(c_0(\s))
 &&=f^{\xx,\xx_1\cdots\xx_s}
(c_0(\s^0))+\sum_{j\ge 1}\big[f^{\xx,\xx_1\cdots\xx_s}(c_0(\s^j))-
f^{\xx,\xx_1\cdots\xx_s}
(c_0(\s^{j-1}))\big] \defin\nn\\
&&\defin\sum_{j\geq0} f^{\xx,\xx_1\cdots\xx_s}_{(j)}
(\s_{nn^{(j)}(\x)})\lb{3.21}\eea
where $\x$ is the spatial coordinate associated to $\xx$ and $nn^{(j)}(\x)=
nn(\x)\times I_j$, $I_j=[-j,j]\cap\ZZZ$. 
Since $|c_0(\s^j)-c_0(\s^{j-1})|\le c\l^j$ for some $c>0$,
$f^{\xx,\xx_1\cdots\xx_s}_{(j)}$ is bounded by:
\be ||f^{\xx,\;\xx_1\dots\xx_s}_{(j)}||_\io
\le G \frac{(s+1)!}{r_0^{s+1}}c\l^j.
\;\lb{3.21a}\ee

   \subsection{Decay of the Potentials for the Conjugation}

Inserting expansion \pref{3.21} in the definition of the value of a
tree eq. \pref{3.7}, we find:
\be \widehat\Val(\th,c_0(\s))=\prod_{v\in\th}\sum_{j_v\ge 0}
 \big(-\a(v)\big) 
\l^{p_v+\rho_{\a(v)}}
  \frac{f_{(j_v)}^{\xx(v),\;\xx(v_1)\dots\;
  \xx(v_{s_v})}}{s_v!}
 (\t^{p(v)}\s_{nn^{(j_v)}(\x(v))})\;,\lb{3.22}\ee
where we recall that $\r_{\a(v)}=(1+\a(v))/2$, $p(v)=\sum_{w\ge v}\a(v)(
p_v+1-\r_{\a(v)})$ and $nn^{(j)}(\x)=nn(\x)\times I_j$, $I_j=[-j,j]\cap \ZZZ$.
The above expression can be seen as a sum over the values of a new
kind of trees, identical to the ones described in section. \ref{ss3.1}, but
with a new label $j_v\in\NNN$ attached to each node. Let $\TT_k(\xx)$
be the set of these new trees of order $k$ contributing to $\d
h^\xx_{(k)}$, \ie $\th\in\TT_k(\xx)$ is a tree with $k$ branches and
$k$ nodes (the root is not a node) with the following labels attached
to the nodes $v\in\th$: $\x(v)\in V$, $p_v\in\NNN$, $j_v\in\NNN$ and
$\a(v)\in\{-1,+1\}$.

Given $\th\in\TT_k(\xx)$, its value is given by:
\be \Val\ \th(\s)=\prod_{v\in\th}\big(-\a(v)\big) 
\l^{p_v+\rho_{\a(v)}}
  \frac{f_{(j_v)}^{\xx(v),\;\xx(v_1)\dots\;
  \xx(v_{s_v})}}{s_v!}
 (\t^{p(v)}\s_{nn^{(j_v)}(\x(v))})\;.\lb{3.23a}\ee
so that we have
\be \d h_{(k)}^\xx(c_0(\s))=\sum_{\th\in\TT_k(\xx)}
\Val\ \th(\s)
\mbox{\ \ \ \ \ with \ }
||\Val\ \th||_\io\le
\Big(\frac{c e G}{\l r_0^2}\Big)^k\prod_{v\in\th}\l^{j_v+p_v +1}
\;.\lb{3.23}\ee
where we have used eq.\pref{3.21a}. We can now define the support
$X(\th)\subset V\times\ZZZ$ of a tree $\th\in\TT_k(\xx)$, as the
support of the spin variables on which $\Val\,\th$ depends in a non
trivial way, plus a {\it center} $(\x,0)$. More precisely:
\be X(\th)\defin \Big\{(\x,0)\Big\}\cup
\bigcup_{v\in\th}\CC\Big(\x(v),p(v),j_v\Big)\;,\qquad
{\hbox{\rm where}}\qquad
\CC\Big(\x(v),p(v),j_v\Big)\defin\bigcup_{\h\in nn(\x(v))}
\bigcup_{|i|\le j_v}(\h, p(v)+i)\;.\lb{3.24}\ee
namely $\CC(\x,p,j)$ is a cylinder centered in $(\x,p)$, with the
spatial base equal to the set of nearest neighbors of $\x$ and with
height equal to $2j$. Then $X(\th)$ is the union of $(\x,0)$ and of
cylinders of this kind, one for each node $v$ of the tree.  The point
$(\x,0)$ has the role of center of $X(\th)$ and is added to $X(\th)$
for later convenience (note in fact that $\Val\ \th(\s)$ could not
depend on $\s_{(\x,0)}$).
 
Given a set $X\subset\ZZZ^{d+1}$ we can partition it in a natural way
as a union of time-like segments. More precisely given $\x\in\ZZZ^d$
let $T_\x=\{(\x,i)\in\ZZZ^{d+1}\,|\, i\in\ZZZ\}$. The intersection
between $T_\x$ and $X$ can be uniquely partitioned as a union of
$n_\x$ maximal connected segments. The collection of all this segments
forms a partition of $X$ in $n_X$ time-like segments
$\{R_i(X)\}_{i=1,\ldots, n_X}$. Let now $r_i$ the center of
$R_i(X)$. If $Y$ is a subset of $\ZZZ^{d+1}$ we call tree distance of
$Y$, $d_t(Y)$, the length of the minimal tree connection all the
points of $Y$. Finally let $d_c(X)$ the tree distance of the set
$\{r_i\}_{i=1,\ldots, n_X}$.

>From the previous bound on the value of a tree
$\th\in\TT_k(\xx)$ eq. \pref{3.23} can be interpreted as the
tree distance decay of the contribution of order $k$ to $\d h$. Indeed: 
\be |\e|^k ||\Val\,\th||_\io
\le \Big(\frac{ce\,G|\e|^{1/2}}{\l r_0^2}\Big)^k
\left[\l^{d_c(X(\th))}|\e|^{
n_{X(\th)}/2}\prod_{i=1}^{n_{X(\th)}}\l^{|R_i(X(\th))|}\right]^{1\over 2d+1}
\;,\lb{3.26}\ee
where:
\begin{enumerate}
\item the factor $\l^{d_c(X(\th))}$ comes from $\prod_{v\in\th}
\l^{1+p_v}$; in fact $p_v$ is the displacement in the time direction 
of the cylinder associated to the node $v$
w.r.t. the one associated to the node $v'$ immediately following $v$,
and $1$ is their maximum displacement in spatial direction, so that
$\sum_v(1+p_v)\ge d_c(X(\th))$;

\item we used that $n_X\le (2d+1)k$ in order to bound $|\e|^{k/2}$
with $|\e|^{n_X\over 2(2d+1)}$;

\item the factor $\prod_{i=1}^{n_{X(\th)}}\l^{|R_i(X(\th))|}$ 
comes from $\prod_{v\in\th}\l^{j_v}$;

\item the global power $1/(2d+1)$ in \pref{3.26} comes from the size
of the base of each cylinder, namely we used the fact that the number
$n_X$ of segments is less than $2d+1$ times the number of cylinders in
$X(\th)$.
\end{enumerate}

Collecting together all the trees $\th$ which have support $X(\th)=X$
for a given $X$, we get:
\be\d h_\e^\xx(c_0(\s))=\sum_{X\ni (\x,0)}\d h_X^\xx
(\s_X) \mbox{\ \ \ \  with \ \ }
 \d h_X^\xx(\s_X)\defin\sum_{k\ge 1}\e^k
\sum_{\th\in\TT_k(\xx)}^{X(\th)\=X}\Val\ \th(\s)\;.\lb{3.25}\ee
So, using the bound \pref{3.26} for $|\e|$ small enough, $\g_0={1\over
2(2d+1)}$, $\k_0=-2\g_0\log\l$, $\n_0=|\e|^{\g_0}$ and a suitable
$c>0$ we get
\be ||\d h^\xx_X||_\io\le c e^{-\k_0 d_c(X)}\n_0^{n_X}\prod_{i=1}^{n_X}e^{-\k_0
|R_i(X)|}\;,\lb{3.27}\ee
namely $\d h_X^\xx$ decays exponentially with the tree distance of $X$.

    \subsection{SRB Potentials and Their Decay}\lb{ssL}

Proceeding as above for the function $\L^\x(c_0(\s))$ we obtain that 
we can write it as:
$$
\L^\x(c_0(\s))\defin\sum_{X\subset(V_N\times\zzz)}
\phi_X^{(\x,0)}(\s_X)
$$
where by construction $\phi_X^{(\x,0)}$ is different from 0 only if
$(\x,0)\in X$. The function  $\phi_X^{(\x,0)}$ is again given by a
tree expansion analogous to that in eq. \pref{3.25}. Moreover we will
set 
$$
\phi_X^{(\x,j)}(\s_X)\defin\phi_{\t^{-j}X}^{(\x,0)}(\s_X).
$$

We can define:
$$
 \phi_X(\s_X)\defin\sum_{(\x,j)\in X}\phi_X^{(\x,j)}(\s_X),
$$
so that we formally obtain eq.\pref{2.11}, namely, given 
$I_T=[-T/2,T/2]\cap\ZZZ$ ($T$ even) and calling $\L=V_N\times I_T$,
$$\sum_{(\x,i)\in \L}
 \L^\x\big(c_0(\t^i\s)\big)
 -
 \sum_{X\cap\L\not =\emptyset}
 \phi_X(\sigma_X)=O(\dpr\L)\;,$$
where $\dpr\L$ is the boundary of $\L$ and the correction can be exactly 
computed from the definitions above.

Note the potential $\phi_X(\s_X)$ is invariant under time and space
translations (respectively for the definition of $\phi^{(\x,i)}_X$ and
for the periodic boundary conditions), namely:
\be
\phi_X(\s_X)=\phi_{\r^\x \t^j X}(\s_X)
\ \ \ \ \mbox{for any}\ \ \ \ (\x,j)\in V_N\times\ZZZ 
.\ee
Moreover it can be bounded by:
\be ||\phi_X||_\io\le c e^{-\k_1 d_c(X)}\n_1^{n_X}\prod_{i=1}^{n_X}e^{-\k_1
|R_i(X)|}\;,\lb{3.29}\ee
for suitable $c,\g_1,\k_1>0$ and $\n_1=|\e|^{\g_1}$.

\section{Analyticity of SRB measure}\lb{ss3.4}\setcounter{equation}{0}

In the previous sections, we wrote the SRB measure as a Gibbs measure
with translationally invariant potentials $\phi_X$, decaying as in
\pref{3.29}, and with hard core interaction in time direction.
Moreover the potential $\phi_X$ is analytic in $\e$ in a small disc in
$\CCC$ around the origin (independent of $N$).  A well known technique
to show analyticity of the Gibbs measure w.r.t. $\e$ is the so called
{\it cluster expansion}.

If $\L=V_N\times I_T$, with $I_T\defin[-T/2, T/2]\cap\ZZZ$ 
for some even $T\in\NNN$, we call $\L_a= V_N\times I_{T+2a}$. Given
a {\it boundary condition} $\bar\s\in\{1,\ldots,n\}^{\zzz^{d+1}}_C$,
we define the {\it pressure} $P_\L$ as 
\be P_\L\defin |\L|^{-1}\log\sum_{\s}e^{-\sum_{X\cap\L\not =
\emptyset}\phi_X(\s_X)}\;.\lb{3.30}\ee
where the sum is over all the $\s$ that coincide with $\s_\L$ on $\L$,
to $\bar\s$ on $\L^c_a$ and with $\Sigma(\s_{\x,T/2},\bar\s_{\x,T/2+a})$
in the space remaining. It
is well known that the pressure $P_\L$ can be considered as the
generating functional for the Gibbs states. From its analyticity our
main theorem will follow easily, as we will see in section 6.4.

  \subsection{Decimation}

In presence of hard cores we cannot proceed  
in the standard way (Mayer's expansion), since the standard proof,
see \cite{GMM73}, 
requires weakness of the original interactions. 
We can overcome this obstacle by a {\it decimation}, see \cite{CO81},
namely considering the statistical system on scales larger than 
the length of decorrelation of the hard core. 

  \subsubsection{Decimated Lattice $\L_D$}
 
For each $\x\in V_N$, we divide the time interval $I_T^\x
\defin\{\x\}\times I_T$ into an alternating sequence of blocks,
called of ``$B$-type'' and ``$H$-type'',
$B_\x^{(0)},H_\x^{(0)},B_\x^{(1)},H_\x^{(1)},\ldots,
B_\x^{(\ell-1)},H_\x^{(\ell-1)},B_\x^{(\ell)}$, containing a number of
spins respectively equal to $b=1$ and $h=h_0a-1$, with $h_0\in\NNN$ to
be chosen later. For this reason we choose the number of points in $I_T^\x$
to be $|I_T^\x|=\ell h_0 a+1$, namely $T=\ell h_0 a$.\\
  
\\{\bf Remark} The choice $b=1$ is special for the present case, in
which the unperturbed potential is vanishing. In general one could
treat with the same technique the case in which the unperturbed
potential is order one, with a sufficiently fast decay of the tails,
and in that case $b$ should be chosen suitably large, see \cite{CO81}.
Such a case arises, for instance, when the unperturbed system is the
product of non linear Anosov maps on $\TTT^2$, namely in the case
treated in Appendix \ref{a2}.  The present discussion could be easily
adapted to cover that case.\\

Let $\b_\x^{(i)}\defin\s_{(\x,-T/2+ih_0a)}$,$\;\x\in V_N$, 
$\;i=0,\ldots, \ell$, 
be the spin in the block $B_\x^{(i)}$ and 
$\h_\x^{(i)}\defin\{\s_{(\x,p)}\}_{(\x,p)\in H_\x^{(i)}}$, 
$\x\in V_N$, $i=0,\ldots,\ell-1$,
be the collection of spins belonging to the block $H_\x^{(i)}$; it
will be regarded as a sequence of $h$ $\b$ spins: $\h^{(i)}_\x=
\big(\b_1(\h^{(i)}_\x),\ldots,\b_h(\h^{(i)}_\x))$. 
The lattice obtained considering the $H$ and $B$ blocks as points:
\be \L_D\defin\{B^{(p)}_\x, H^{(q)}_\x\}^{q=0, \ldots,
\ell-1}_{\x\in V_N,\, p=0,\ldots,\ell}\ee
will be called the {\it decimated lattice}; on $\L_D$ the distances
will be computed by thinking of it as having its sites spaced by 1
also in the time direction.

If $X\subset\L$, $Y(X)$ will denote the corresponding subset in  
$\L_D$, namely the smaller subset $Y\subset\L_D$ such that the union
of the $B$ and $H$--blocks in $Y$ contains the set $X$. Defining  
$\Phi_Y(\b_{Y},\h_{Y})\defin \sum_{X: Y(X)=Y}\phi_X(\s_X)$, eq.\pref{3.30} 
can be rewritten as
\bea
&&P_\L={1\over |\L|}\log\sum_{\b_{\L}}\sum_{\h_{\L}}
e^{-\sum_{Y\subset\L_D}\Phi_Y(\b_{Y},\h_{Y})}
\prod_{\x\in V}
\prod_{i=0}^{\ell-1} Z(\b_\x^{(i)},\h_\x^{(i)},
\b_\x^{(i+1)})\;,\nn\\
&&\hbox{\rm where}\qquad Z(\b,\h,\b')=C_{\b\b_1(\h)}
C_{\b_1(\h)\b_2(\h)}\cdots C_{\b_{h-1}(\h)\b_h(\h)}
C_{\b_h(\h)\b'}\;.\lb{6.2}\eea
Observe that, from eq. \pref{3.29}, if $Y$ do not coincide with a single
$H$--block, $\Phi_Y$ satisfies a qualitatively equivalent bound:
\be ||\Phi_Y||_\io\le c\, e^{-\tilde\k d_c(Y)}\widetilde\n^{n_Y}
\prod_{i=1}^{n_Y}e^{-\tilde\k|R_i(Y)|}\virg Y\not =H_\x^{(i)}\;,\lb{6.2a}\ee
for some $c,\widetilde\k,\widetilde\g>0$ and $\widetilde\n=
|\e|^{\tilde\g}$. Whereas if $Y=H_\x^{(i)}$ for some $\x\in V$ 
and some $i=0,\ldots,\ell-1$, we have
$||\Phi_Y||_\io\le h\widetilde\n$.

 \subsubsection{Averaging over many degree of freedom: 
               the Perron-Frobenius's theorem}

Decimation is a {\it renormalization group} technique, consisting in
summing first on the $H$-type spins, thus getting an effective
statistical system for the $B$-blocks: the idea is that if the
$B$-blocks are sufficiently far apart, after the averaging of the
$\h$'s, the $\b$'s should be {\it almost independent}, as if there
were only small interactions among them. The technical tool we shall
use to prove rigorously that the effective interactions between
the $\b$'s are small is Perron-Frobenius theorem.

Let $Z(\b,\b')$ be defined, with a little abuse of notation, as:
\be Z(\b,\b')\,\defin\,\sum_\h Z(\b,\h,\b')=C^{ah_0}_{\b\b'}\lb{6.4}\ee
Observe that $1\le C^{a}_{\s\s'}\le q^{a}$. Since $C^{a}$ has strictly
positive entries, we can apply the Perron-Frobenius theorem and obtain
that $C^{a}$ and its transpose $C^{a,T}$ admit a non degenerate
eigenvalue $l>0$ with eigenvectors $\p$ and
$\p^*$ respectively such that $\p_\s,\ \p^*_\s>0$ for any $\s=1,...\ q$, and
$\sum_\s\p^*_\s\p_\s=1$. The eigenvalue $l$ is maximal in the spectrum of 
$C^a$; namely, if we define $P$ as the projection 
matrix $P_{\s\s'}=\d_{\s\s'}-\p_\s\p^*_{\s'}$, we have:
\be ||(l^{-1}C^{a})^kP\o||_\io
\le c_\a e^{-\a k}||\o||_\io\;,\lb{6.5a}\ee
for any $\o\in\RRR^q$ and with 
\be\a\defin-\log(1-[\min(C^{a}_{\s\s''}/C^{a}_{\s\s'})]^2)
\ge q^{-2a}\;.\lb{6.5}\ee
As a consequence,
\bea Z(\b,\b')&&=C^{ah_0}_{\b\b'}=\sum_\s C^{ah_0}_{\b\s}(\p_\s\p^*_{\b'}
+P_{\s\b'})=\nn\\
&&=l^{h_0}\p_\b\p^*_{\b'}
\Bigl[1+{(l^{-h_0}C^{ah_0}P)_{\b\b'}\over\p_\b\p^*_{\b'}}\Bigr]\defin
l^{h_0}\p_\b\p^*_{\b'}e^{-I(\b,\b')}\;,\lb{6.6}\eea
with $I(\b,\b')=O(e^{-h_0 q^{-2a}})$. It is now clear that taking
$h_0$ big enough we can make the two body potential $I(\b,\b')$ as
small as needed.

Using eq. \pref{6.6}, introducing a new effective potential $W$ 
including the contributions from $\Phi$ and $I$, defining
\be \prod_{i=0}^\ell e^{-U^{(i)}(\b^{(i)}_\x)}\defin
\prod_{i=0}^{\ell-1}\p_{\b^{(i)}_\x}\p^*_{\b^{(i+1)}_\x}\lb{6.7}\ee 
and using $\lim_{\L\to\io}|\L|^{-1}\log\prod_{\x,i}\sum_{\b^{(i)}_\x}
e^{-U^{(i)}
(\b_\x^{(i)})}=0$ (as it follows from the normalization condition 
$\sum_\s\p^*_\s\p_\s=1$), we can rewrite $P_\L$ as 
\bea && P_\L={1\over a}\log l+
{1\over |\L|}\log\sum_{\b_{\L}}\sum_{\h_{\L}}
m(\b_{\L},\h_{\L})\prod_{Y\subset\lis\L}
e^{-W_Y(\b_{Y},\h_{Y})}
\nn\\&& \hbox{\rm with}\quad
m(\b_{\L},\h_{\L})\defin
\prod_{\x\in V}\prod_{i=0}^\ell {e^{-U^{(i)}(\b_\x^{(i)})}\over \sum_{\b_\x}
e^{-U^{(i)}(\b_\x)}}
\prod_{i=0}^{\ell-1}
{Z(\b^{(i)}_\x,\h^{(i)}_\x,\b^{(i+1)}_\x)\over Z(\b^{(i)}_\x,\b^{(i+1)}_\x)}
\lb{6.8}\eea
where $m(\b_{\L},\h_{\L})$ is a probability density. Observe that, if
one chooses $h_0\simeq -\log\widetilde\n$ (so that both
$h\widetilde\n$ and $I(\b,\b')$ are small), the new interaction
$W$ satisfies a bound similar the one of $\Phi$:
\be||W_Y||_\io\le c e^{-\lis{\k}d_c(Y)}
\lis{\n}^{n_{Y}}\prod_{i=1}^{n_{Y}}e^{-\bar{\k}|R_{i}(Y)|}\virg
\forall Y\subset\L_D
\;,\lb{6.3a}\ee
for some $c,\lis\k,\lis\g>0$, $\lis\n=|\e|^{\lis\g}$.

      \subsection{Mayer's Expansion and Polymer Lattice Gas}

We shall now expand the small potential appearing in the 
expression for $P_\L$, via a {\it Mayer's expansion}, 
obtaining the pressure for $\e=0$ plus a correction.

It will be convenient to   
collect together the contributions of the potentials whose supports
have the same {\it closure}, in the following sense: for a set formed
by a unique point $H_\x^{(i)}\in\L_D$ we define its {\it closure} as
$\lis {(H_\x^{(i)}})\,\defin\, (B_\x^{(i)}, H_\x^{(i)}, B_\x^{(i+1)})$
while for a set formed by a unique point $B_\x^{(i)}\in\L_D$ we define
$\lis {(B_\x^{(i)})}\,\defin\, (B_\x^{(i)})$; finally for
$Y\subset\L_D$ we define its closure as $\overline Y\,\defin\,
\cup_{G\in Y}\overline {(G)}$.

We say that a collection $\CC=\{Y_m\}_{m=1}^n$ of sets
$Y_i\subset\ZZZ^{d+1}$ (think to them as {\it molecules}) is {\it
connected} if, given a couple $(Y_{\mbox{\scriptsize
in}},Y_{\mbox{\scriptsize fin}}) \in\CC\times\CC$, it is possible to
find $\{Y_{m_j}\}_{j=1}^p$, such that $\lis Y_{\mbox{\scriptsize
in}}\cap \lis Y_{m_1}\not =\emptyset$, $\lis Y_{m_i}\cap\lis
Y_{m_{i+1}}\not =\emptyset$ and $\lis Y_{m_p}\cap \lis
Y_{\mbox{\scriptsize fin}} \not =\emptyset$.

Writing $e^{-W_Y(\b_{Y},\h_{Y})}$ as 
the value for $\e=0$ plus the correction, namely 
$1+\Big(e^{-W_Y(\b_{Y},\h_{Y})}-1\Big)$, expanding the product 
over $Y\subset \L_D$ and collecting together the connected
components, we can rewrite eq.\pref{6.8} as:
\be P_\L-{1\over a}\log l
={1\over |\L|}\log\sum_{\b_{\L}}\sum_{\h_{\L}}
m(\b_{\L},\h_{\L})\sum_{\G\subset \L_D}\Y(\G)\prod_{\g\in \G}
\r(\g|\b_{\g},\h_{\g})\;,\lb{6.9}\ee
where:
\begin{enumerate}
\item{} $\g$ is a subset of $\L_D$, to be called in the following 
{\it polymer} (they are, indeed, the union of a connected 
collection of molecules);
\item{} $\G$ is a collection of polymers: $\G=(\g_1,\ldots,\g_n)$, $n\ge 1$ 
and $\G\subset\L_D$ means that $\g\subset\L_D$, $\forall \g\in\G$; 
\item{} $\Y(\G)$ is the function equal 
to $1$ if $\g\cap\g'=\emptyset$ for every
$\g,\g'\in X$ with $\g \not =\g'$ and $0$ otherwise;
\item{} $\r(\g | \b_{\g},\h_{\g})$, $\g\subset\L_D$ is defined as
\be \r(\g|\b_{\g},\h_{\g})\defin \ 
\sum_{q\ge 1}{1\over q!}\sum_{Y_1,\ldots,
Y_q \atop \cup_i\lis Y_i=\g}^*\prod_{i=1}^q\left(e^{-W_{Y_i}
(\b_{Y_i},\h_{Y_i})}-1\right)\lb{6.10}\ee
where the $*$ on the sum means that $Y_1,\ldots,Y_q$ is a connected 
collection of subsets of $\L_D$;
\item{} the term corresponding 
to $\G=\emptyset$ must be interpreted as equal to 1.
\end{enumerate}

The key observation is that, thanks to the above definition of
closure, in \pref{6.9} we can sum over $\h$ spins before summing over
the $\b$ spins. After doing this the measure $m(\b_{\L},\h_{\L})$
factorizes, \ie
\bea P_\L-{1\over a}\log l
&&={1\over |\L|}\log\sum_{\G\subset\L_D}\Y(\G)
\prod_{\g\in \G}\left[\sum_{\b_{\g}}\sum_{\h_{\g}}
m(\b_{\g},\h_{\g})
\r(\g|\b_{\g},\h_{\g})\right]\defin\nn\\
&&\defin{1\over |\L|}\log\sum_{\G\subset\L_D}\Y(\G)
\prod_{\g\in \G}\r(\g)
\;,\lb{6.11}\eea
Namely we have rewritten $P_\L$ as the pressure for $\e=0$
plus a correction having the form of the pressure of a ``polymer
lattice gas'', with activities $\r(\g)$ and hard core potentials
$\Y(\G)$.

    \subsection{Cluster Expansion and Its Convergence}

A standard argument, exposed for instance in \cite{GMM73},
\cite{PS99} or \cite{GBG03},
leads to
\be P_\L-{1\over a}\log l=
{1\over |\L|}\log\sum_{\G\subset\L_D}\Y(\G)\r(\G)=
{1\over |\L|}\sum_{\G\subset\L_D}\Y^T(\G)\r(\G)\;,\lb{6.13}\ee
where $\Y^T$ is the Mayer function, defined as
\be \Y^T(\g_1,\ldots,\g_n)\defin\left\{
\begin{array}{ll}
\sum_{g\in\GG(n)}\prod\limits_{(i,j)\in g}f(\g_i,\g_j) & \mbox{if}\ \  n>1\\
                  1                                    & \mbox{if}\ \  n=1.
\end{array}\right.\;,\lb{3.32}\ee
where $\GG(n)$ is the set of connected graphs which can be drawn on
$n$ vertices labeled $1,\ldots,n$ by connecting with links couples of
distinct vertices; the function $f(\g_i,\g_j)$ is equal to $1$ if
$\g_i\cap\g_j \not =\emptyset$ and $0$ otherwise. By construction,
$\Y^T(\G)$ is different from zero only if $\G$ is a connected
collection of polymers. Observe that $\G$ could contain many copies of
the same $\g$. More precisely here $\G$ represents a function from the
subsets of $\L_D$ to $\NNN$ (and we can think $\G(\g)$ as representing 
the number of copies of $\g$)
such that $\sum_{\g\subset\L_D}\G(\g)\leq
\infty$. 

A bound for $\r(\g)$ can be obtained as follows:
\be |\r(\g)|\le ||\r(\g|\cdot,\cdot)||_\io\le\sum_{p\ge 1}{1\over p!}
\sum_{Y_i,\ \cup_i\lis Y_i=\g}^*\prod_{i=1}^p ||W_{Y_i}||_\io e^{
||W_{Y_i}||_\io}\;,\lb{3.33}\ee
Using the bound \pref{6.3a} (and that, if $\cup_{i=1}^p\lis Y_i=\g$,
one has $\sum_{i=1}^p||W_{Y_i}||\leq c\lis\n|\g|$)  we find:
\be |\r(\g)|\le e^{c\lis\n|\g|}\sum_{p\ge 1}{1\over p!}
\sum_{Y_i,\ \cup_i\lis Y_i=\g}^*\prod_{i=1}^pc e^{-\lis{\k}d_c(Y_i)}
\lis{\n}^{n_{Y_i}}
\prod_{j=1}^{n_{Y_i}}e^{-\lis{\k}|R_{j}(Y_i)|} \lb{3.33a}\;.\ee
We can now use the connectedness constraint on the sum in order to extract 
a factor exponentially small in the size of $\g$. Indeed, 
if $\cup_{i=1}^p\lis Y_i=\g$, one has both 
$\sum_{i=1}^p d_c(Y_i)\geq d_c(Y_i)$ and $\sum_{i=1}^p n_{Y_i}\geq n_\g$. 
After extracting such a 
factor we can relax the constraints on the sum, so that:
\be |\r(\g)|\le c e^{c\lis\n|\g|}\left[ e^{-{\lis{\k}\over 2}d_c(\g)}
\lis{\n}^{n_{\g}/2}
\prod_{i=1}^{n_{\g}}e^{-{\lis{\k}\over 2}|R_{i}(\g)|} \right]
\sum_{p\ge 1}{1\over p!}\Big(
\sum_{\lis Y\subset\g}e^{-{\lis{\k}\over 2}d_c(Y)}
\lis{\n}^{n_{Y}/2}
\prod_{j=1}^{n_{Y}}e^{-{\lis{\k}\over 2}|R(Y)|}\Big)^p\;.\lb{3.33b}\ee
It is easy to see that the last sum is bounded by $c|\g|\lis\n^{1/4}$,
so that:
\be |\r(\g)|\le c e^{c\bar\n|\g|}\left[ e^{-{\bar{\k}\over 2}d_c(\g)}
\bar{\n}^{n_{\g}/2}
\prod_{i=1}^{n_{\g}}e^{-{\bar{\k}\over 2}|R_{i}(\g)|} \right]
\sum_{p\ge 1}{1\over p!}(c|\g|\bar\n^{1/4})^p\le
c e^{-\k'd_c(\g)}(\n')^{n_{\g}}
\prod_{i=1}^{n_{\g}}e^{-\k'|R_{i}(\g)|} \;,\lb{6.14}\ee
for some $c,\k',\g'>0$ and $\n'=|\e|^{\g'}$. Using the preceding bound 
we can easily prove that:
\be\sup_{x\in\zzzz^{d+1}}\sum_{\d(\g)
\ge r}^{\g\ni x}|\r(\g)|\le c(\n')^{1/2}e^{-({\k'/2})r}\;.\lb{3.33c}\ee
where $\d(\g)$ is the diameter of the polymer $\g$.
A standard theorem, proved for instance in \cite{GMM73}\cite{GBG03}, states 
that, if $\r(\g)$ satisfies \pref{3.33c}, then
\be \sup_{x\in\L_D}\sum_{\d(\G)\ge r}^{\G\ni x}\Y^T(\G)|\r(\G)|\le c(\n')^{
1/4}e^{-(\k'/4)r}\;.\lb{3.34}\ee
This implies that, varying $\L$, $P_\L$ is a uniformly
convergent sequence of analytic functions in a domain independent from
$\L$.  The limit, still analytic in the same domain (thanks to
Vitali's convergence theorem), is independent of the
way the thermodynamic limit is performed (\ie one can send the time
side of $\L$ to $\io$ either before the spatial side is sent to $\io$
or together with it), thanks to the exponentially fast convergence of
the sequence, implied by \pref{3.34}.  For the same reason, the limit is
also independent of the choice of boundary conditions and, because of
translational invariance, it is equal to:
\be P\defin\lim_{|\L|\to\io}P_\L=
{1\over a}\log l+{2\over h_0 a}
\sum_{\G\subset\zzz^{d+1}}^{\G\ni (0,0)}{\Y^T(\G)\r(\G)\over |\G|}
\;,\lb{3.35}\ee
where $|\G|\defin |\cup_{\g\in\G}\g|$ and $2/(h_0 a)=\lim_{|\L|\to\io}|\L_D|/
|\L|$.

\subsection{Analyticity of the Mean Values}\lb{ss6.4}

The analyticity for the mean value of an analytic local observable
$\OO(\psi_{V})$ (depending on the variables in the finite set
$V\subset\ZZZ^d$) is an easy corollary of the previous result.

We first observe that
$\m^{SRB}(\OO)=\lim_{N,T\to\infty}{1\over|V_N||I_T|}
\sum_{(\x,i)\in(V_N\times I_T)} \m^{SRB}(\OO\circ \r^\x\circ S_\e^i)$. 
This is true thanks to the
time and space translation invariance of $\m^{SRB}$. Moreover it is possible to
decompose $\OO$ as
$$
 \OO \big(h_\e(c_0(\s))_{V}\big)
 =
 \sum_{X\cap (V\times\{0\})=\emptyset}\OO_X^{(0,0)}(\s_X)
$$
This can be done expanding $\OO (h_\e)$ in power of $\psi$,
using the representation of $h_\e$ given in section 3 and 5 and
collecting the terms with the same support. Moreover we will set:
$$
\OO^{(\x,j)}_{X}(\s_X)\defin\OO^{(0,0)}_{\rho^{-\x}\t^{-j}X}(\s_X)
$$
and 
$$
\OO_X(\s_X)\defin\sum_{(\x,j)\atop \rho^\x V\times \{j\}\cap X
\not=\emptyset} \OO^{(\x,j)}_X(\s_X)
$$
It is easy to realize that $\OO_X$ is invariant 
under space and time translations,
and satisfies
\be||\OO_X||_\io
\le c_V\n^{n_X}e^{-\k d_c(X)}
\prod_{i=1}^{n_X}e^{-\k|R_i(X)|}\;,\lb{3.36}\ee
for some $\k,\g>0$, $\n=|\e|^{\g}$ and some constant $c_{V}>0$ which
depends on the size of $V$. Setting $\L=V_N\times I_T$, the
thermodynamic limit of the mean value of $\OO(\psi_{V})$ can be
written as:
\be \m^{\SRB}(\OO)=\lim_{\L\to\io}{1\over |\L|}\dpr_\z\log{\sum_{\s_\L}
e^{-\sum_{X\cap\L\not =\emptyset}[\phi_X(\s_X)-\z\OO_X(\s_X)]}
\over\sum_{\s_\L}
e^{-\sum_{X\cap\L\not =\emptyset}\phi_X(\s_X)}}
\Biggr|_{\z=0}\defin\partial_\z P_\OO(\z) \;.\lb{3.36b}\ee
Via a new cluster expansion we find: 
\be \m^{\SRB}(\OO)=\lim_{\L\to\io}{1\over |\L|}\dpr_\z
\sum_{\G\cap\L_D\not=\emptyset}\Y^T(\G)\big(\r^\z(\G)-\r(\G)\big)
\Bigr|_{\z=0}\;,\lb{3.37}\ee
where $\r^\z(\g)$ are the activities corresponding to the potential
$\phi_X-\z\OO_X$.  For $|\z|$ small enough, the potential
$\phi_X-\z\OO_X$ satisfies the same bounds of $\phi_X$ so that
$\sum_{\G\cap\L_D\not=\emptyset} \Y^T(\G)\big(\r^\z(\G)-\r(\G)\big)$
is a uniformly convergent sequence of functions, analytic in $\e$ and
$\z$ in the product of two small discs.  This implies that
$\m^{\SRB}(\OO)$ is analytic in $\e$ and given by:
\be \m^{\SRB}(\OO)={2\over h_0 a}\sum_{\G\subset\zzz^{d+1}}^{\G\ni(0,0)}
{\Y^T(\G)
\over |\G|}\dpr_\z\big(\r^\z(\G)-\r(\G)\big)\Bigr|_{\z=0}\;.\lb{3.38}\ee 

\newpage

\appendix

\section{Unperturbed Nonlinear Dynamics}\lb{a2}\setcounter{equation}{0}

The result about analyticity can be extended to the case in which
the unperturbed dynamic is made up of independent {\it nonlinear} 
analytic Anosov systems $s_0:\TTT^2\to\TTT^2$. We suppose that there
exist $v_\pm(\psi)$ and $\l_\pm(\psi)$ such that:

\be \big(Ds_0\ v_+\big)(\phi)=
 \lambda_+(\phi)v_+\big(s_0(\phi)\big)
 \virg \big(Ds_0\ v_-\big)(\phi)
 =\lambda_-(\phi)v_-\big(s_0(\phi)\big)\;,\lb{7.1}\ee
with $\phi\in\TTT^2$, $v_\pm(\phi)$ and $\l_\pm(\phi)$ H\"older
continuous and $|\lambda_+(\phi)|^{-1}$, $|\lambda_-(\phi)|$
$\leq\lambda<1$. Then we consider again a perturbation $f(\psi)$ 
on $\TT_N$ analytic in $\psi$. 
Observe however that in this case the most naive example of 
perturbation, $f(\psi)\defin f^+(\psi)v_+(\psi)$, 
with $f^+(\psi)$ analytic, is {\it no longer} an
analytic perturbation.

 \subsection{Conjugation}

The constitutive equation for $\d h_\e$, lifted on $\RRR^{2 V_N}$,
is:
\be
 S_0\big(h_\e(\psi)\big)
 +
 \e f\big(h_\e(\psi)\big)
 =
 S_0(\psi)+\d h_\e\big(S_0(\psi)\big)
.\ee
In order to exploit the hyperbolicity, it is convenient to arrange 
the terms as follows:
\be
 \big(DS_0\ \d h_\e\big)(\psi)
 -
 \d h_\e\big(S_0(\psi)\big)
 =
 -
 \e f\big(\psi+\d h_\e(\psi)\big)
 -
 \big[
 S_0\big(h_\e(\psi)\big)-S_0(\psi)-\big(DS_0\ \d h_\e\big)(\psi)\big]
.\ee
Define $$
 f^{\ ,\xx_1\cdots\ \xx_s}(\psi)
 \defin
 \frac{\dpr^s}{\dpr\z_1,...\ \dpr\z_s}
 f\Big(\psi+\z_1 w^{(\x_1)}_{0,\a_1}(\psi)+...\ 
\z_s w^{(\x_s)}_{0,\a_s}(\psi)\Big)
 \Big|_{\z_1=...\ \z_s=0}
$$
whereas
$$f^{\ ,\xx_1\cdots\ \xx_s}(\psi)
\defin \sum_\xx f^{\xx,\xx_1\cdots \xx_s}(\psi) w^{(\x)}_{0,\a}
\big(S_0(\psi)\big)
\ \ \ \ \mbox{and}\ \ \ \ S_0^{\ ,\xx_1\cdots \xx_s}(\psi)
\defin \sum_\xx S_0^{\xx,\xx_1\cdots \xx_s}(\psi) w^{(\x)}_{0,\a}
\big(S_0(\psi)\big);$$
%
Writing $\d h_\e(\psi)\defin\sum_{\xx}\d
h^\xx_\e(\psi)w^{(\x)}_{0,\a}(\psi),$ and $\big(DS_0\
w^{(\x)}_{0,\a})(\psi) \defin \sum_{\yy}S_0^{\yy,\xx}(\psi)
w^{(\h)}_{0,\b}\big(S_0(\psi)\big)$, with
$S_0^{\yy,\xx}(\psi)=\l_\a(\psi_\x) \d_{\xx,\yy}$, 
we get:
\be
 \l_\a \d h^\xx_\e(\psi)-\d h^\xx_\e\big(S_0(\psi)\big)
 =
 -
 \e\sum_{s\geq0}
 \Big(\frac{f^{\xx,\xx_1\cdots \xx_s}}{s!}
 \d h_\e^{\xx_1}\cdots\d h_\e^{\xx_s}\Big)(\psi)
 -
 \sum_{s\ge 2}
 \Big(\frac{S_0^{\xx,\xx_1...\;\xx_s}}{s!}
 \d h^{\xx_1}_\e...\;\d h_\e^{\xx_s}\Big)(\psi)
,\ee
Finally, the recursive equation for 
the Taylor coefficients of $\d h_\e^{\x^+}(\psi)$ is:
\bea \d h^{\x^+}_{(k+1)}(\psi)
=&& - \sum_{p\geq0}\Big(
 \prod_{m=0}^p\lambda_+^{-1}\big(s_0^m(\psi_\x)\big)\Big)\sum_{s\ge 0}
\sum_{k_1+\cdots+k_s=k\atop k_i\ge 1}
\Big(\frac{f^{\x^+,\xx_1\cdots \xx_s}}{s!}
\d h_{(k_1)}^{\xx_1}\cdots\d h_{(k_s)}^{\xx_s}\Big)\big(S_0^p(\psi)\big)+
\lb{7.4}\\ 
&& +\sum_{p\geq0}\Big(
 \prod_{m=0}^p\lambda_+^{-1}\big(s_0^m(\psi_\x)\big)\Big)\sum_{s\ge 2}
\sum_{k_1+\cdots+k_s=k+1\atop k_i\ge 1}
\Big(\frac{S_0^{\x^+,\xx_1\cdots\xx_s}}
{s!}\d h_{(k_1)}^{\xx_1}
\cdots \d h_{(k_s)}^{\xx_s}\Big)\big(S_0^p(\psi)\big)\;.\nn\eea
A similar equation holds for $\xx=\x^-$.

>From now on, the construction of the conjugation function goes on as
in the linear case with similar considerations.  We have only to take
in account the fact that a tree of order $k$ (w.r.t. $\e$) does not
necessarily have $k$ branches, because of the term on the last line of
\pref{7.4} (to be called a vertex of type $0$).  Since the number of
lines entering a vertex of type $0$ is $\ge 2$, one can easily prove
that the number $b_k$ of branches of a tree of order $k$ is bounded by
$k\le b_k\le 2k-1$, so that nothing qualitatively changes in the
bounds and the proof of analyticity of $\d h_\e$ proceed as in
sections sections \ref{s3} and \ref{ss3.4}.

     \subsection{Unstable Direction}

The perturbed unstable direction in the point $h_\e(\psi)$ 
is given by the equation:
\be
 \Big(DS_\e\ w^{(\x)}_{\e,+}\Big)\big(h_\e(\psi)\big)
 =
 w^{(\h)}_{\e,+}\Big(h_\e\big(S_0(\psi)\big)\Big)L^{\h\x}(\psi)
.\ee
Setting $w^{(\x)}_{\e,+}\big(h_\e(\psi)\big)\defin v^{(\x)}_\e(\psi)$, 
it is convenient to rearrange the terms of the equation in the following way:
\bea
 \big(DS_0\ v^{(\x)}_\e\big)(\psi)
 -
 \l_+(\psi_\x)v^{(\x)}_e\big(S_0(\psi)\big)
 =
&&\d L^{\h\x}(\psi)v^{(\h)}_\e\big(S_0(\psi)\big)\nn\\
&&-\e\Big(Df\Big)\big(h_\e(\psi)\big)v^{(\x)}_\e(\psi) 
 -
 \Big[DS_0\big(h_\e(\psi)\big)-DS_0(\psi)\Big]v^{(\x)}_\e(\psi)
.\lb{1.87}\eea
Defining $v^{(\h)}_\e(\psi)=\sum_\xx
V^{(\h)}_{\e,\xx}(\psi)w^{(\x)}_{0,\a}(\psi)$, and using again the
considerations of section \ref{ss3.2}, we finally get 
%
\bea 
  \l_\a(\psi_\r)V_{\e,\xx}^{(\r)}(\psi)
  - 
  \l_+(\psi_\r)V_{\e,\xx}^{(\r)}\big(S_0(\psi)\big)
  =
&&+
  \delta L^{\z\r}(\psi)V_{\e,\xx}^{(\z)}\big(S_0(\psi)\big)
  \nn\\
&&-
  \e\sum_{s\geq0}
  \Big(\frac{f^{\xx,\yy\xx_1\cdots \xx_s}}{s!}
  V_{\e,\yy}^{(\r)}
  \d h_\e^{\xx_1}\cdots\d h_\e^{\xx_s}\Big)(\psi)\nn\\
&&-
  \sum_{s\ge 1}
  \Big(\frac{S_0^{\xx\yy,\xx_1...\;\xx_s}}{s!}
  V_{\e,\yy}^{(\r)}\d h^{\xx_1}_\e...\;\d h_\e^{\xx_s}\Big)(\psi)
  \; \nn
\eea 
(with $\big(DS_0^{\ \ ,\xx_1...\;\xx_s}w^{(\x)}_a\big)(\psi)
\defin S_0^{\xx\yy,\xx_1...\;\xx_s}(\psi)w^{(\h)}_b(S_0\psi)$).
Again, because of the third term in the r.h.s.  of eq.\pref{1.87} ,
the number of branches of a tree appearing in the construction of $\d
V$ and $\d L$ is greater (in general) than the order of the tree
itself. This is not a problem, since one can easily realize that,
again, the number $b_k$ of branches of a tree of order $k$ is such
that $k\le b_k\le 2k$.

     \subsection{SRB Interactions}

Following the proof in section \ref{ss3.2} and, proceeding as in
sections \ref{ss3.3} and 6 one proves analyticity of SRB
distribution. In fact the only (slight) difference in the construction
of SRB potentials is in the telescopic cutting necessary to represent
$h$, $L$, $V$ and $\L$ as sums of local functions of spin
variables. Notice that now each tree node is associated to the product
of a node function $f_v(\psi)$ (\eg in the case of a tree contributing
to $\d h$, $f_v$ can be a derivative of $f$ or a derivative of $S_0$,
see \pref{7.4}) times a product of local Lyapunov exponents, like the
factor $\prod_{m=0}^p\lambda_+^{-1}\big(s_0^m(\psi_\x)
\big)\defin\Theta_+(p,s_0^{p(v)}(\psi_\x))$ in \pref{7.4}; the
analogous expression appearing in a vertex with $\a(v)=-$ will be
denoted by $\Theta_-(p,s_0^{p(v)}\psi_\x)$.  So the total node
function associated to a vertex $v$ will be now of the form
\be F^v(p(v),S_0^{p(v)}(\psi))\defin\Theta_{\a(v)}
(p(v),s_0^{p(v)}(\psi_{\x_v}))f_v(S_0^{p(v)}(\psi))\;,\lb{7.5a}\ee
where $v'$ is the vertex immediately following $v$. The telescopic
expansion \pref{3.21} has to be done {\it separately} for each of the
factor in the above equation ($\l(\psi)$ is h\"older continuous),
getting in the end potentials with the same kind of decay rate. 
The bounds are not qualitatively changed and the subsequent analysis of
section \ref{ss3.3} follows so that, by suitably modifying the decimation
procedure, analyticity of SRB measure can be proved. We point out that
a main difference in the proof of convergence of the cluster expansion
is that now the unperturbed potentials are not vanishing, but have
support only on time like segments $I\subset\ZZZ$, and are
exponentially decaying with the diameter of $I$. For this reason one
cannot proceed exactly as in section 6. The standard way to treat this
problem, see \cite{CO81}, is to choose a length $r$ such that the
unperturbed interactions on sets $I$, $\mbox {diam}(I)>r$, are small
enough for the cluster expansion. Then one fix the size of the
$B$-blocks $b=r$, and the size of the $H$-blocks, $h$, such that the
Perron-Frobenius theorem is true for the {\it reduced partition function}
$Z_r(\b_\x^{(p)},\h_\x^{(p)},\b_\x^{(p+1)})$, in which only the
interaction on sets $I\subset (B_\x^{(p)}\cup H_\x^{(p)}\cup
B_\x^{(p+1)})$, $\mbox {diam}(I)\leq r$ are taken in account.
 
\section{Green-Kubo Formula and Large Deviation}\lb{s4}\setcounter{equation}{0}

In this section we deal with an application. We introduce 
the {\it local phase space contraction rate} on a volume $V_0\subset V_N$ 
averaged on a time $T_0$, given by
\be \h_{\L_0}(\psi)\,\defin\, {1\over
|\L_0|}\sum_{j\in I_{T_0}}\log\Big| \det (D S_\e)_{V_0}\big(S_\e^j(\psi)\big)
\Big|\;,\ee
with $\L_0=V_0\times I_0$ and $I_0\defin[-T_0/2,T_0/2]\cap\ZZZ$.  We
prove a Green-Kubo formula for $\h_{\L_0}$, from which it will come
out that generically its mean value $\h_+$ is strictly negative.
Furthermore, we can show the large fluctuations of
$\h_{\L_0}$ around $\h_+$ satisfy a large deviation
principle, namely they are asymptotically described by a strictly
convex {\it free energy functional} $F(\h)$: it can be
obtained as the Legendre transform of the generating functional
$P(\z)=P_{\h_{\L_0}}(\z)$, see eq. \pref{3.36b}. 

For the rest of the appendix the SRB interaction will be called
$\{\phi^+_X\}_{X\subset\zzz^{d+1}}$, to remind that they are derived
from the unstable restriction of $DS_\e$. \\ \\ {\bf Theorem B1}: {\it
Given $S_\e$ such that $\h_+<0$,
\begin{enumerate}
\item  $P(\z)$ is analytic  and strictly convex in $\z$, for 
$|\e|< \e_0$, $|\z|\le 1$, with $\e_0$ small enough;
\item  the Green-Kubo formula is valid:
\be
  \partial^2_\e P'(0)|_{\e=0}=-\frac{1}{2}\partial^2_\e P''(0)|_{\e=0}.
\ee
\end{enumerate}}
\\
{\bf Theorem B2}: {\it Given $S_\e$ such that $\h_+<0$,
\begin{enumerate}
\item the free energy $F(\h)$ is analytic in $\h$, for 
$|\e|< \e_0$, and $\h\in[P'(-1),P'(1)]$;
\item if $[a,b]\subset[P'(-1),P'(1)]$, then 
\be\lim_{|\L_0|\to\io}{1\over |\L_0|}\log \m^{\SRB}\big(\h_{\L_0}\in[a,b]\big)=
\max_{\h\in[a,b]}-\D F(\h,\h_+)\;,\ee
with $\D F(\h,\h_+)\defin F(\h)-F(\h_+)$.
\end{enumerate}}

\subsection{Local Phase Space Contraction Rate}\lb{ss4.1}

Repeating the construction of SRB potentials leading to \pref{2.11}, 
we set:
\be
\h_{\L_0}(h_\e(c_0(\s)))\defin{1\over|\L_0|}
\sum_{X\subset \zzz^{d+1}}^{X\cap\L_0\neq0}\phi_X(\s_X)
\;,\lb{4.1}\ee
for a suitable potential $\phi_X$, satisfying
\be ||\phi_X||_\io\le c e^{-\k d_c(X)}\n^{n_X}\prod_{i=1}^{n_X}
e^{-\k|R_i(X)|}\;,\lb{4.1b}\ee
for some $c,\k,\g>0$ and $\n=|\e|^\g$.
>From the invariance under time translations of the SRB measure, 
we have
\bea \h_+&\defin&
\lim_{|V_0|\to\io}
{1\over |V_0|}\m^{\SRB}\big(\log\Big|\det (D S_\e)_{V_0}\Big|
\big)=\lim_{|\L_0|\to\io}\m^{SRB}\big(\h_{\L_0}\big)=\nn\\ 
&=&\lim_{|\L_0|\to\io}{1\over |\L_0|}{\sum}_{X\cap \L_0\not=\emptyset}
\m^{\SRB}(\phi_X)=\lb{4.4}\\
&=&\lim_{|\L_0|\to\io}\lim_{|\L|\to\io}{1\over |\L_0|}\left.\dpr_\z\log{
\sum_{\s_{\L}}e^{-\sum_{X\cap{\L}\not=\emptyset}\phi^+_X(\s_X)+\z
\sum_{X\cap{\L_0}\not=\emptyset}\phi_X(\s_X)}\over
\sum_{\s_{\L}}e^{-\sum_{X\cap\L\not=\emptyset}\phi^+_X(\s_X)}}\right|_{\z=0}
\;.\nn\eea
It is easy to show the last expression is equal to the one with 
the summations over $X\cap\L\not = \emptyset$ and $X\cap\L_0\not = \emptyset$ replaced by 
$X\subset\L_0$ and without
the limit in $\L$ (since the correction is only a border effect; 
or simply using again the cluster expansion developed in section \ref{ss6.4}). 
In this way, defining the {\it generating function} $P(\z)$,  as
\be P(\z)\defin\lim_{|\L|\to\io}{1\over |\L|}
\log{\sum_{\s_\L}e^{-\sum_{X\subset{\L}}\left(\phi^+_X-\z\phi_X
\right)(\s_X)}\over
\sum_{\s_\L}e^{-\sum_{X\subset\L}\phi^+_X(\s_X)}}\;.\lb{4.2}\ee
 we finally get:
\be\h_+=P'(0)\;.\lb{4.5}\ee
Analyticity is achieved by cluster expansion (we do not need $\z$ small,
but we can take, say, $|\z|\leq1$, since $\{\phi_X\}_{X}$ are O($\e$)).
\subsection{Green-Kubo Formula}\lb{ss4.2}

Consider the case in which $s_0$ is the Arnold's cat map defined by \pref{1}. 

Using the definition of pressure \pref{4.2} 
and the fast convergence properties of the cluster expansion of 
$P(\z)$, we find:
\be P(\z)=\lim_{|\L|\to\io}{1\over |\L|}\log{\m^{\SRB}_{N,0}\Bigl(
e^{-\sum_{j\in I_T}\log|\det L\circ S_0^j|
+\z\sum_{j\in I_T}\log|\det DS_\e\circ h_\e\circ S_0^j|}\Bigr)\over
\m_{N,0}^{\SRB}\Bigl(e^{-\sum_{j\in I_T}\log|\det L\circ S_0^j|}
\Bigr)}\;,\lb{4.6}\ee
where:
\begin{enumerate}

\item the matrix $L=\LL\circ h_\e$ was introduced in section \ref{ss2.2} above;
\item $\m^{\SRB}_{N,0}$ is the unperturbed SRB measure: if $\OO(\psi)$ is 
a local h\"older continuous observable, it is defined as 
\be \m^{\SRB}_{N,0}(\OO)=\lim_{|\L|\to\io}{\sum_{\s_{\L}}\OO
(c_0(\s_{\L}|\hat\s_{\L^c}))\over\sum_{\s_{\L}} 1}\lb{4.7}\ee
and, independently of the boundary conditions, it is equal to
the Lebesgue measure.
\end{enumerate}

Defining $U_\z$ as
\be U_\z=\log|\det L|-
\z\log|\det S_0^{-1}\circ DS_\e\circ h_\e|\;,\lb{4.9}\ee
and using that $\m^{\SRB}_{N,0}$ is the Lebesgue measure on $\TT_N$,
we find:
\be P(\z)=\lim_{|\L|\to\io}{1\over |\L|}
\log{\int\der\psi\
e^{-\sum_{j\in I_T}U_\z(S_0^j\psi)}\over
\int\der\psi\
e^{-\sum_{j\in I_T}U_0(S_0^j\psi)}}
\;,\lb{4.10}\ee
so that $P'(0)$ is equal to
\be P'(0)=\lim_{|\L|\to\io}{1\over |\L|}\sum_{j\in I_T}
{\int\der\psi\log|\det S_0^{-1} DS_\e(h_\e(S_0^j\psi))|
\ e^{-\sum_{j\in I_T}U_0(S_0^j\psi)}\over
\int\der\psi\
e^{-\sum_{j\in I_T}U_0(S_0^j\psi)}}\;.\lb{4.11}\ee
Since $P'(0)|_{\e=0}$ is trivially $=0$, we can try to see if $\dpr_\e 
P'(0)|_{\e=0}$ is different from zero (if it were, $P'(0)$ would be different 
from zero for $\e\not=0$ small enough). Recalling that $f(\psi)$ is 
the perturbing function and $f^\x(\psi)$ is its projection
on the $\x$--th site, we get:
\bea \dpr_\e P'(0)|_{\e=0}&&=\lim_{|\L|\to\io}{1\over |\L|}\sum_{j\in I_T 
}\int{\der\psi\over (2\p)^{2|V_N|}}\Tr[S_0^{-1}D f(S_0^j\psi)]=\nn\\
&&=\lim_{|\L|\to\io}{1\over |\L|}\sum_{j\in I_T}
\sum_{\a=\pm\atop \x\in V_N}
\int{\der\psi\over(2\p)^{2|V_N|}}\l^{\a}f^{\x^\a,\x^\a}(S_0^j\psi)
\;.\lb{4.12}\eea
Since $f$ is periodic we have $\dpr_\e P'(0)|_{\e=0}=0$. 

A straightforward calculation shows that
\bea &&
{1\over 2}\dpr^2_\e P'(0)|_{\e=0}=
\lim_{|\L|\to\io}{1\over |\L|}\sum_{j\in I_T}
\int{\der\psi\over(2\p)^{2|V_N|}}\Bigl\{
\Tr[S_0^{-1}D^2 f(S_0^j\psi)\d h_{(1)}(S_0^j\psi)]-\nn\\
&&-{1\over 2}\Tr[(S_0^{-1}D f(S_0^j\psi))^2]-\sum_{j'\in I_T}\Tr
(S_0^{-1}D f(S_0^j\psi))\Tr^{(u)}(S_0^{-1}D f(S^{j'}_0\psi))\Bigr\}
\;,\lb{4.13}\eea
where $\Tr^{(u)}$ is the trace restricted to the (unperturbed) unstable
manifold. The preceding expression can be rewritten in a more convenient way.
Using the explicit expression of $\d h_{(1)}$ eq.\pref{3.5}, 
and defining $A_0=\cup_{\x\in nn(0)}nn(\x)$,
we find that the first term in eq.\pref{4.13} is equal to
\bea &&\sum_{\a_i=\pm\atop
|\x|\le 1}\sum_{p\ge 0}
\int{\der\psi_{A_0}\over(2\p)^{2|A_0|}}\l^{\a_1}
f^{0^{\a_1},0^{\a_1}\x^{\a_2}}(\psi)(-\a_2)\l^{p+\r_{\a_2}}
f^{\x^{\a_2}}(S_0^{\a_2(p+1-\r_{\a_2})}\psi)=\nn\\&&=\sum_{\a_i=\pm\atop
|\x|\le 1}\sum_{p\ge 0}
\int{\der\psi_{A_0}\over(2\p)^{2|A_0|}}\l^{\a_1}
f^{0^{\a_1},0^{\a_1}}(\psi)\a_2\l^{\a_2}
f^{\x^{\a_2},\x^{\a_2}}(S_0^{\a_2(p+1-\r_{\a_2})}\psi)\;.\lb{4.14}\eea
Integrating by parts, we see that 
the sum of the second and third term in eq. \pref{4.13} is equal to:
\be-{1\over 2}\sum_{\a_i=\pm\atop
|\x|\le 1}\int{\der\psi_{A_0}\over(2\p)^{2|A_0|}}
\l^{\a_1}f^{0^{\a_1},0^{\a_1}}(\psi)\Bigl[\l^{\a_2}
f^{\x^{\a_2},\x^{\a_2}}(\psi)+\l\sum_{p\in\zzzz}
f^{\x^+,\x^+}(S_0^p\psi)\Bigr]\;.\lb{4.15}\ee
Combining the three contributions, we finally find:
\bea\dpr^2_\e P'(0)|_{\e=0}&&=-\sum_{\a_i=\pm\atop|\x|\le 1}\sum_{p\in\zzzz}
\int{\der\psi_{A_0}\over(2\p)^{2|A_0|}}\l^{\a_1}
f^{0^{\a_1},0^{\a_1}}(\psi)\l^{\a_2}
f^{\x^{\a_2},\x^{\a_2}}(S_0^p\psi)=\nn\\
&&=-\lim_{|\L|\to\io}{1\over |\L|}
\int{\der\psi\over(2\p)^{2|V_N|}}\Bigl(\sum_{j\in I_T}
\Tr[S_0^{-1} Df(S_0^j\psi)]\Bigr)^2=-{1\over 2}\dpr^2_\e P''(0)|_{\e=0}
\;,\lb{4.16}\eea
that is the expected Green-Kubo relation (see \cite{GR97}).

>From eq.\pref{4.16} and eq.\pref{4.5}, 
we see that, for $\e$ small enough, $\h_+$ is
negative and, generically, strictly negative (the condition for $f$ to be 
{\it generic} is just that the first line in eq.\pref{4.16} is different 
from 0).

Let us now compute eq.\pref{4.16} in a special case, 
essentially the simpler possible. Let 
\be f^{\x^+}(\psi)=
\sum_{\h\in nn(\x)}\sin(\psi_\x^1-\psi_\h^1)\virg
f^{\x^-}(\psi)=0\;.\lb{4.16aa}\ee
Substituting such choice in eq.\pref{4.16}, we find:
\be\dpr^2_\e P'(0)|_{\e=0}=
-2\sum_{|\x|= 1}\int{\der\psi_0\over(2\p)^2}{\der\psi_\x
\over(2\p)^2}
\l^2\cos^2(\psi_0^1-\psi_\x^1)(v_+\cdot\hat e_1)^2=
-{2d\over 1+\l^{-2}}\;,\lb{4.17}\ee
where $\hat e_1=(1,0)$ and we used that $v_+=({1\over \sqrt{1+\l^2}},-
{\l\over \sqrt{1+\l^2}})$.

So, choosing $\e\in\RRR$ small enough and different from zero, 
$\h_+=P'(0)=-{d\over 1+\l^{-2}}\e^2+O(\e^3)<0$. Furthermore, if $\z\in\RRR$
has modulus smaller than $1$, $P(\z)$ is strictly convex 
(since ${1\over 2}P''(0)=-P'(0)>0$ and $P(\z)$
is analytic for $|\z|\le 1$ and $\e$ small enough).

\subsection{Large deviations}\lb{s5}

In the present section we shall prove 
a large deviations property for $\h_{\L_0}$.
We will follow the classical strategy set up in \cite{Si}, \cite{GLM02}
(in particular we will refer to the formulas in section 5 of the latter).
The proof below will hold in the case $\h_+<0$, namely in the generic case or,
to be definite, in the case the perturbation is chosen as in eq.\pref{4.16aa}.

Thanks to the convexity of $P(\z)$, given $\h\in[P'(-1),P'(1)]$,
there exist a unique point $Z(\h)\in [-1,1]$ such that 
$P'\big(Z(\h)\big)\equiv\h$; considering such a point $\h$ and its 
neighbor of radius $\d$, $I_\d(\h)$, such that 
$I_\d(\h)\subset[P'(-1),P'(1)]$, from 
the ``large deviation property III'', section 5 of \cite{GLM02},
we get:
\be
 \m^{\SRB}\big(\h_+\in I_\d(\h)\big)
 =
 O(1)e^{O(\d|\L_0|)}e^{O(|\dpr \L_0|)}
\exp \Big\{\big[P\big(Z(\h)\big)-P(0)-Z(\h)\h\big]|\L_0|\Big\}\;.\nn\ee
In our case $P(0)=0$. Still for $\h\in[P'(-1),P'(1)]$,
we define the free energy $F(\h)$ as the Laplace transform of the 
generating function $P(\z)$:
\be F(\h)\defin \max_\z\{\z\h- P(\z)\}
 =Z(\h)\h-P\big(Z(\h)\big)
 \;;\lb{b1}\ee
therefore, for $I_\d(\h)\subset[P'(-1),P'(1)]$
\be\m^{\SRB}\Big(\h_+\in I_\d(\h)\Big)=
O(1)e^{O(\d|\L_0|)}e^{O(|\dpr \L_0|)}\exp\Big\{-|\L_0|\D F(\h,\h_+)\Big\}\;;
\lb{5.6}\ee
where $\D F(\h,\h_+)\defin F(\h)-F(\h_+)$ (indeed $F(\h_+)=-P(0)=0$).

Finally, if 
$[a,b]\subset[P'(-1),P'(1)]$,
it is suitable to  take
$\d_{\L_0}=|\L_0|^{-\b}$, $0<\b<1$, and divide the interval $[a,b]$ in 
$|b-a||\L_0|^{\b}$
identical disjoint subintervals centered in 
$\h_n\defin a+(n-1/2)\d_{\L_0}$, we find:
\bea
 \m^{\SRB}\Big(\h_{\L_0}\in[a,b]\Big)
 &&=
 \sum_{n=1}^{|b-a||\L_0|^{\b}}
 \m^{\SRB}\Big(\h_+\in I_{\d_{\L_0}}(\h_n)\Big)\\
 &&=
 O(1)
 |\L_0|^{\b}e^{O(|\L_0|^{1-\b})}e^{O(|\dpr\L_0|)}
 \exp\Big\{|\L_0|\max_{\h\in[a,b]}\big[-\D F(\h,\h_+)\big]\Big\}\;;\nn\eea
namely the result in the second theorem.

\section*{ \hspace*{7.8cm}\normalsize Acknowledgment}

{\small\sl P.F. thanks Benedetto Scoppola for his help in
understanding the techniques related to the Cluster Expansion. The
authors are indebted to Giovanni Gallavotti for many suggestions and
discussions during the preparation of this work.}

\end{document}